%% file: main.tex
\documentclass[10pt,twocolumn]{article}
\usepackage[margin=0.75in]{geometry}
\usepackage{times}

\usepackage{amsmath, amssymb, amsfonts}
\usepackage{algorithmic}
\usepackage{array}
\usepackage[caption=false,font=normalsize]{subfig}
\usepackage{textcomp}
\usepackage{graphicx}
\usepackage{cite}
\usepackage{xcolor}
\usepackage{multirow}
\usepackage{booktabs}
\usepackage{tabularx}
\usepackage{pifont}
\usepackage[ruled,vlined]{algorithm2e}
\usepackage{verbatim}
\usepackage{url}
\usepackage{siunitx}
\usepackage{colortbl}
\usepackage{hyperref}
\usepackage{amsthm}

\newtheorem{definition}{Definition}

\newtheorem{remark}{Remark}

\SetCommentSty{mycommfont}
\SetKwBlock{Client}{$\mathcal{C}$:}{}
\SetKwBlock{Server}{$\mathcal{S}$:}{}
\SetKwBlock{TrustedParty}{$\mathcal{TP}$:}{}
\SetKwBlock{SetUp}{Setup:}{}
\SetKwBlock{VerificationStage}{Verification Stage:}{}
\SetKwBlock{RegisterStage}{Registration Stage:}{}
\SetKwInput{KwInput}{Input}
\SetKwInput{KwOutput}{Output}

\newcommand{\cmark}{\ding{51}}
\newcommand{\xmark}{\ding{55}}

\title{Bidirectional Biometric Authentication Using Transciphering and (T)FHE}

\author{
Joon Soo Yoo\textsuperscript{1}, Tae Min Ahn\textsuperscript{2}, and Ji Won Yoon\textsuperscript{1} \\[0.5em]
\small \textsuperscript{1}School of Cybersecurity, Korea University, Seoul 02841, Republic of Korea \\
\small \textsuperscript{2}Korean Testing Certification Institute (KTC), Gyeonggi-do 15809, Republic of Korea \\
}
\date{} % hides the date

\begin{document}
\maketitle

\input{Sections/00_abstract}

\input{Sections/01_introduction}
\input{Sections/02_pre}

\input{Sections/03_problem}

\input{Sections/04_our_model}

\input{Sections/05_network_efficiency}

\input{Sections/06_key_management}

\input{Sections/07_security_privacy}

\input{Sections/08_evaluation}
\input{Sections/09_conclusion}

\bibliographystyle{plain} 
% \bibliography{bib}

\input{main.bbl}
\end{document}

%% file: Sections/00_abstract.tex
\begin{abstract}
Biometric authentication systems pose privacy risks, as leaked templates such as iris or fingerprints can lead to security breaches. Fully Homomorphic Encryption (FHE) enables secure encrypted evaluation, but its deployment is hindered by large ciphertexts, high key overhead, and limited trust models. We propose the Bidirectional Transciphering Framework (BTF), combining FHE, transciphering, and a non-colluding trusted party to enable efficient and privacy-preserving biometric authentication. The key architectural innovation is the introduction of a trusted party that assists in evaluation and key management, along with a double encryption mechanism to preserve the FHE trust model, where client data remains private. BTF addresses three core deployment challenges: reducing the size of returned FHE ciphertexts, preventing clients from falsely reporting successful authentication, and enabling scalable, centralized FHE key management. We implement BTF using TFHE and the Trivium cipher, and evaluate it on iris-based biometric data. Our results show up to a 121$\times$ reduction in transmission size compared to standard FHE models, demonstrating practical scalability and deployment potential.
\end{abstract}

%% file: Sections/01_introduction.tex
\section{Introduction}

Authentication serves as the first line of defense, ensuring that only legitimate users are authorized to access resources, data, or systems. As cloud services expand rapidly, especially with the rise of online banking, e-commerce, and IoT devices, weak authentication mechanisms can lead to significant security issues. Therefore, the need for a robust authentication system is critical for verifying user identity and protecting sensitive information.

The current biometric authentication system, particularly for online authentication without the use of a Trusted Execution Environment (TEE)~\cite{tee-1}, requires the server to store either the raw or transformed biometric data of the user. Storing raw biometric data presents a significant privacy risk, as this sensitive information becomes a prime target for attackers if the server is compromised. To mitigate this risk, technologies such as cancelable biometrics~\cite{cancel-1} and biometric cryptosystems (BCS)~\cite{bcs-1} transform the biometric data into a non-reversible format. While these transformed data types do not expose the raw biometric information directly in the event of a breach, they are not entirely risk-free. If the transformed data or the associated helper data in BCS is not sufficiently secured, attackers may still exploit this information to impersonate the user. Although these systems are designed to minimize such risks through robust cryptographic techniques, vulnerabilities can still arise depending on the implementation.

Fully Homomorphic Encryption (FHE)~\cite{fhe-intro, fhe-survey} is highly suited for privacy-sensitive scenarios because it allows servers to perform computations on encrypted data without needing to decrypt it, unlike traditional models. This ensures that biometric data, once encrypted on the client side, remains encrypted during transmission, storage, and evaluation (matching). Furthermore, FHE is primarily based on lattice-based cryptography, which is considered resistant to quantum attacks, making it post-quantum secure. Given these advantages, constructing biometric authentication systems using FHE is emerging as a promising trend, offering a higher level of security and resilience against server compromises.

Despite FHE being one of the most secure cryptographic technologies, it still faces several challenges when it comes to practical deployment—particularly in terms of evaluation time and ciphertext size. For instance, in the \textsf{TFHE}~\cite{tfhe-1, tfhe-2} scheme, which is a logic-based FHE approach, a ciphertext can be more than $2,000$ times larger than a single plaintext bit, depending on the security level\footnote{For example, when $n=630$, for a $\lambda = 128$-bit security level}. Similarly, the \textsf{CKKS}~\cite{ckks-1, ckks-2} encryption scheme, which is a popular arithmetic-based scheme, allows SIMD operations that pack multiple plaintexts into a single ciphertext, yet still struggles with large ciphertext sizes. Therefore, the significant ciphertext expansion, compared to symmetric ciphers that do not require such overhead, remains a major obstacle to the practical use of FHE.

\input{Tabs/related_tabs}

Transciphering~\cite{trans-prop, transc-1, transc-2, transc-3, transc-4} is a technique for converting one encryption scheme to another to achieve a balance between security and efficiency. In the context of FHE, transciphering allows the system to leverage the powerful evaluation capabilities of FHE on the server side, while relying on lightweight symmetric encryption during transmission between the client and server.

The key idea is to homomorphically decrypt a stream cipher ciphertext that has itself been encrypted under the FHE scheme. Specifically, the client first encrypts a message~$m$ using a symmetric stream cipher with key~$k$, resulting in a ciphertext $c = \mathsf{E}(m, k)$. Rather than encrypting $m$ directly under FHE, the client sends~$c$ to the server. The server then encrypts~$c$ under the FHE public key $\mathsf{pk}$, producing $\mathsf{Enc}(c, \mathsf{pk})$. It then performs a homomorphic decryption of $c$ to obtain $\mathsf{Enc}(m)$—the message now encrypted under the FHE scheme. This transformation enables the client to benefit from the low overhead of symmetric encryption while offloading the more costly homomorphic operations to the server. Importantly, because the client transmits only a small symmetric ciphertext, the network overhead remains comparable to traditional encrypted communication, avoiding the large transmission costs associated with native FHE ciphertexts.

\subsection{Transciphering Challenges in Authentication Systems}

While it may seem straightforward to naively adapt existing transciphering frameworks to the domain of FHE-based biometric authentication, several practical challenges arise in this setting. Most notably, current transciphering approaches primarily focus on minimizing the transmission overhead from the client to the server. However, they often overlook the size of the ciphertext returned to the client after computation, which can also contribute significantly to the overall network overhead.

Another challenge arises from the trust model typically assumed in FHE-based systems. After the server completes the homomorphic evaluation, the authentication result remains encrypted under the client’s key, meaning only the client can decrypt it. In this setting, the server must rely on the client to report whether authentication was successful. However, this creates a security vulnerability: a malicious client could falsely claim a successful authentication, thereby compromising the integrity of the system. As a result, the traditional FHE model---where only the client possesses the decryption key---can be fundamentally unsuitable for biometric authentication, where the correctness of the result must be verifiable by the server.

Lastly, the server must handle multiple cryptographic keys from each client, which introduces substantial scalability concerns. In particular, the client is required to transmit both a public key for encrypting the stream ciphertext under the FHE scheme and an evaluation key for enabling homomorphic operations. Under our TFHE implementation, the size of the public key is approximately 4.93 MB, while the evaluation key reaches 41.6 MB. These values correspond to a single client. As the number of clients increases, the cumulative overhead of managing and transmitting these large keys can overwhelm the network, significantly impacting the system's scalability and practicality.

\subsection{Our Work}

In this work, we propose a novel approach called the \textsf{Bidirectional Transciphering Framework (BTF)}, which leverages transciphering techniques and FHE to improve the network efficiency of biometric authentication systems. Our goal is to enable the practical deployment of secure, privacy-preserving FHE-based authentication. A key architectural change in our framework, compared to the traditional FHE setting, is the introduction of a trusted party—similar to what is commonly assumed in practical biometric systems. Importantly, the client’s private data remains hidden from both the server and the trusted party, under the assumption that these entities do not collude. Each party holds a portion of the keying material for the doubly encrypted data, and thus, without collusion, neither entity can recover the client’s sensitive information.

Our framework systematically addresses the challenges outlined above. First, in our framework, the FHE-encrypted authentication result generated by the server is not returned directly to the client. Instead, it is sent to the trusted party, which performs the decryption using the FHE secret key. The trusted party then re-encrypts the result using a symmetric stream cipher and sends this lightweight ciphertext to the client. As a result, multi-kilobyte FHE ciphertexts are never transmitted to the client, thereby reducing the network overhead that would otherwise be incurred in traditional FHE-based systems. 

Second, the authentication result is decrypted and verified by the trusted party, preventing the client from falsely claiming successful authentication. 

In addition to result verification, our framework also assigns the trusted party the role of generating and distributing the FHE keys, including the public key and evaluation key, to both the client and the server. Since only the trusted party holds the secret key and performs decryption, a single key pair ($\mathsf{pk}, \mathsf{evk}$) can be safely shared among multiple clients under the assumption that the trusted party does not collude with the server. This eliminates the need for each client to generate and transmit distinct FHE keys, which is required in traditional models and scales poorly.

Our contributions are as follows:

\begin{itemize}
    \item We propose a novel framework for biometric authentication that leverages transciphering to reduce network overhead and address key management challenges, enabling a scalable and practical FHE-based protocol.

    \item We introduce a bidirectional model that eliminates the need to transmit large FHE ciphertexts to the client by offloading result decryption to a trusted party and returning a lightweight stream cipher ciphertext.

    \item We propose a double encryption scheme that protects the symmetric key used for stream cipher encryption against both the server and the trusted party, ensuring privacy even under semi-honest but non-colluding assumptions.

    % \item We centralize FHE key generation and distribution using the trusted party, which enables key reuse across clients and significantly reduces the communication overhead typically associated with large evaluation keys and public keys.

    \item We implement and evaluate our system using the TFHE and Trivium schemes, demonstrating up to $121\times$ reduction in client-to-server transmission size compared to baseline FHE models, along with practical scalability across multiple clients.
\end{itemize}

\subsection{Related Works}

Privacy-preserving biometric authentication has been extensively studied, particularly in the domains of biometric cryptosystems~\cite{bcs-1, bcs-2}, cancellable biometrics~\cite{cancel-1, cancel-2}, and biometric authentication leveraging HE~\cite{multi-bio}. For a thorough review in these areas, the reader is referred to the comprehensive survey by~\cite{comprehen-bio}. In this section, however, we focus on discussing the most relevant works that incorporate transciphering and FHE within the context of biometric authentication. A comparison of these works is presented in Table~\ref{tab:bio_comparison}.

Homomorphic transciphering was first introduced by Naehrig~\cite{trans-prop}, and since then, several subsequent works have emerged. Balenbois et al.~\cite{transc-1} present a framework for transciphering using the \textsf{TFHE} scheme in conjunction with \textsf{TFHE}-friendly ciphers, namely \textsf{Trivium}~\cite{trivium} and \textsf{Kreyvium}~\cite{kreyvium}. Their work proposes a single-server model for the transciphering framework, evaluating various encoding methods to design a homomorphic decryption circuit with the goal of optimizing time performance. 

Cho et al.~\cite{transc-2} provide a transciphering framework based on the \textsf{CKKS} scheme, introducing a CKKS-friendly cipher called HERA. Their work presents an optimized homomorphic decryption circuit tailored to the CKKS scheme. Both works~\cite{transc-1, transc-2} assume a single-server model where only the valid user possesses the secret key. However, neither work addresses the challenge of handling the returning FHE ciphertext, as the secret key remains solely with the client.

Zhou et al.~\cite{bio-partial} employ a functional encryption (FE) scheme~\cite{func-enc} to construct a privacy-preserving biometric authentication system. The central idea is that the user's encrypted biometric features are compared, and the result is returned in plaintext using FE, after which the server makes an authentication decision based on the result. However, their work does not address the issue of transciphering during the transmission phase, which leads to network overhead.

Barrero et al.~\cite{multi-bio} present a biometric authentication system based on the Paillier cryptosystem~\cite{paillier}, an additive HE scheme grounded in the decisional composite residuosity assumption. The proposed system employs a two-server architecture: one server stores the encrypted biometric template, while the user computes the similarity score based on newly extracted features from the probed fingerprint, encrypts the result, and sends it to an authentication server. The authentication server decrypts the result and notifies both the client and the database server of the authentication outcome. Although the work does not specifically address network overhead caused by Paillier ciphertexts, it should be noted that the overhead is relatively small compared to that of FHE ciphertexts.

%% file: Tabs/related_tabs.tex
\begin{table*}[!htb]
\centering
\footnotesize
\caption{Comparison of Approaches in Biometric Authentication. ($\mathsf{SC}$ represents stream cipher)}
\begin{tabularx}{\textwidth}{|c|c|c|c|c|c|>{\centering\arraybackslash}X|}
\hline
\textbf{Reference} & \textbf{Single Server} & \textbf{Transcipher} & \textbf{Return Size} $\downarrow$ & \textbf{Biometric System} & \textbf{Symmetric Cipher} & \textbf{Encryption Scheme} \\ \hline
Balenbois et al.~\cite{transc-1} & \cmark & \cmark & \xmark & \xmark &  \textsf{Trivium} \& \textsf{Kreyvium} & \textsf{TFHE} \\ \hline
Cho et al.~\cite{transc-2} & \cmark & \cmark & \xmark & \xmark & \textsf{HERA} & \textsf{CKKS} \\ \hline
Zhou et al.~\cite{bio-partial} & \cmark & \xmark & \xmark & \cmark & \textsf{N/A} & \scriptsize{\textsf{Functional Encryption}} \\ \hline
Barrero et al.~\cite{multi-bio} & \xmark & \xmark & \xmark & \cmark & \textsf{N/A} & \textsf{Paillier} \\ \hline
\textbf{Our Work} & \xmark & \cmark & \cmark & \cmark & \textsf{FHE-friendly SC} & \textsf{FHE} \\ \hline
\end{tabularx}
\label{tab:bio_comparison}
\end{table*}

%% file: Sections/02_pre.tex
\section{Preliminaries}

\subsection{Notation}

Bold uppercase letters denote matrices (e.g., $\mathbf{A}$), bold lowercase letters indicate vectors (e.g., $\mathbf{a}$), and italic letters represent scalars (e.g., $a$). For binary representations of scalars, we use bracket notation (e.g., $a[i]$), where $a[i]$ is the $i$-th bit of scalar $a$. The FHE scheme is denoted by $\mathsf{Enc}$, while $\mathsf{E}$ refers to the symmetric stream cipher ($\mathsf{SC}$) scheme. For a positive integer $k$, $[k] = \{1, 2, \ldots, k\}$ denotes the index set. 
We use superscripts such as $^{\mathsf{FHE}}$ to indicate homomorphic operations. For example, $\oplus^{\mathsf{FHE}}$ denotes the XOR operation performed on ciphertexts under the FHE scheme.

\subsection{Homomorphic Encryption (HE)}

HE enables computations to be performed directly on encrypted data. A special type of HE, called Fully Homomorphic Encryption (FHE), supports both addition and multiplication operations with an arbitrary number of operations. One of the most prominent underlying problems that FHE leverages is the Learning With Errors (LWE)~\cite{lwe-1, lwe-2} problem, which is considered resistant to quantum attacks.

Since Gentry's introduction of the bootstrapping technique~\cite{gentry-fhe}, which reduces ciphertext noise by homomorphically decrypting noisy ciphertexts, enabling an arbitrary number of operations on encrypted data, FHE has been extensively researched. Subsequent work has introduced various FHE schemes, including GSW-based~\cite{gsw, tfhe-1, fhew} and BGV-based~\cite{bgv, ckks-1} schemes. GSW-based schemes are optimized for implementing Boolean logic gates, while BGV-based schemes efficiently support arithmetic operations such as addition and multiplication.

\noindent \textbf{Torus Fully Homomorphic Encryption (TFHE).} TFHE~\cite{tfhe-1, tfhe-2} is among the most widely used schemes within GSW-based FHE~\cite{gsw}, utilizing different types of ciphertexts for message encryption and bootstrapping over the torus $\mathbb{T} = \mathbb{R} / \mathbb{Z}$, i.e., real numbers modulo 1. 
\input{Sections/Defs/lwe_sample}
\input{Sections/Defs/decisional}

The decisional LWE assumption states that solving the LWE problem—that is, distinguishing an LWE sample from a random sample with a probability better than \(1/2\)—is computationally infeasible for some security parameter \(\lambda\).

\input{Sections/Defs/pubkey}

We present the basic functionality of TFHE, focusing exclusively on the LWE sample-based approach as follows.

\begin{itemize}
    \item \( (\mathsf{pk}, \mathsf{sk}, \mathsf{evk}) \stackrel{\$}{\leftarrow} \mathsf{Enc.KeyGen}(1^{\lambda_{\mathsf{FHE}}}) \). The secret key \(\mathsf{sk} = \mathbf{s}\), public key \(\mathsf{pk}\), and evaluation key \(\mathsf{evk}\) are generated based on the security parameter \( \lambda_{\mathsf{FHE}} \).
    
    \item \( \mathsf{ct} \leftarrow \mathsf{Enc}(m, \mathsf{pk}) \). The message \( m \in \mathbb{B} \) is encrypted using the public key \(\mathsf{pk}\), producing the LWE ciphertext \(\mathsf{ct}\). Specifically, the public key \(\mathsf{pk} = (\mathbf{a}, b)\) is an LWE sample under the secret key \(\mathbf{s}\), and the ciphertext is defined as \(\mathsf{ct} = (\mathbf{a}, b) + (\mathbf{0}, \mathsf{Ecd}(m))\), where \(\mathsf{Ecd}\) is an encoding function in TFHE that maps \(m\) to a torus element \(\mu\), with \(1 \mapsto 1/4\) and \(0 \mapsto 0\).

    \item \( m \leftarrow \mathsf{Dec}(\mathsf{ct}, \mathsf{sk}) \): Given the ciphertext \(\mathsf{ct} = (\mathbf{a}, b)\) and the secret key \(\mathsf{sk} = \mathbf{s}\), compute \( b - \mathbf{a} \cdot \mathbf{s} \) to retrieve the torus element \(\mu\) with some noise \(e\). Then, round \(\mu\) to the nearest encoding value, either \(0\) or \(1/4\), and decode ($\mathsf{E^{-1}}$) it to obtain the original message \( m \).

    \item \( \mathsf{Enc}(f(m_1, m_2), \mathsf{pk}) \leftarrow \mathsf{Eval}(f, \mathsf{ct_1}, \mathsf{ct_2}, \mathsf{evk}) \). The evaluation function \(\mathsf{Eval}\) performs the operation \( f \) on the ciphertexts \(\mathsf{ct_1}\) and \(\mathsf{ct_2}\) using the evaluation key \(\mathsf{evk}\), producing a new ciphertext that corresponds to \( f(m_1, m_2) \).
\end{itemize}

\input{Sections/Defs/rem_1}

\input{Sections/Defs/rem_2}

\subsection{Transciphering}

Transciphering is a technique that converts ciphertexts from one encryption scheme to another, typically from a symmetric encryption scheme to a FHE scheme. This concept was first proposed by Naehrig et al.~\cite{trans-prop}, who introduced transciphering as a potential solution to mitigate network congestion caused by the large ciphertexts in FHE systems.
% ---- our work overview
\begin{figure}[!htb] 
  \centering 
  \includegraphics[width=0.47\textwidth]{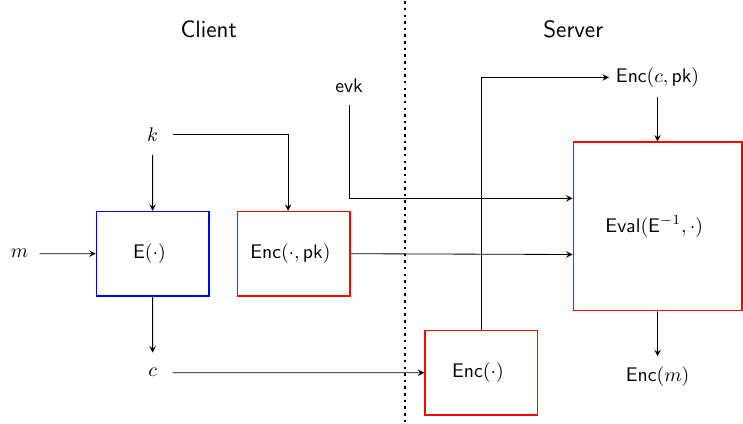}
  \caption{General Transciphering Framework.} 
  \label{fig:transc-general} 
\end{figure}

The transciphering framework operates as follows: the client first encrypts a message \( m \) using a symmetric cipher \( \mathsf{E} \) with key \( k \), producing the ciphertext \( c = \mathsf{E}(m, k) \). The client transmits \( c \) along with the evaluation key \( \mathsf{evk} \) to the server. Upon receipt, the server encrypts \( c \) under the FHE scheme, yielding \( \mathsf{Enc}(c, \mathsf{pk}) \). Next, the server uses \( \mathsf{evk} \) to homomorphically evaluate the decryption function of the symmetric cipher, \( \mathsf{Eval}(\mathsf{E}^{-1}, c) \), resulting in \( \mathsf{Enc}(m) \), the FHE-encrypted form of the original message without revealing the plaintext.

As shown in Fig.~\ref{fig:transc-general}, during transmission, the symmetric ciphertext \( c \) is sent to the server, which significantly reduces network congestion compared to sending \( \mathsf{Enc}(m, \mathsf{pk}) \) directly. At the final stage of transciphering, the server obtains \( \mathsf{Enc}(m, \mathsf{pk}) \), enabling homomorphic evaluation on \( m \) with the privacy-preserving properties of FHE. However, when the server produces a result \( \mathsf{Enc}(r, \mathsf{pk}) \), it cannot decrypt it. Therefore, the FHE ciphertext must be sent back to the client, introducing network congestion at the return phase.

\subsection{FHE-friendly Stream Cipher}

Original works~\cite{aes-fhe, aes-fhe-2} focused on homomorphic implementations of AES; however, these showed poor performance in terms of homomorphic decryption of the AES circuit. This inefficiency is due to AES not being suitable for homomorphic operations, as it involves large multiplicative depth. Subsequent research proposed FHE-friendly ciphers, such as Rasta~\cite{rasta}, Dasta~\cite{dasta}, Trivium~\cite{trivium}, Kreyvium~\cite{kreyvium}, FiLIP~\cite{filip}, and Pasta~\cite{pasta}, which are defined over \( \mathbb{F}_2 \) (i.e., the plaintext values are binary). These ciphers are designed to be lightweight and efficient for FHE evaluation, enabling faster homomorphic decryption of stream cipher.

As a representative example, Trivium operates using three shift registers of different lengths. After random selection of the key \( \mathbf{k} \) and the initial vector \( \mathbf{IV} \), the internal state undergoes 1,152 iterations of updating through the following recurrence relations over \( \mathbb{F}_2 \):
\begin{align*}
v_i &= z_{i-111} + z_{i-110} \cdot z_{i-109} + z_{i-66} + v_{i-69}, \\
w_i &= v_{i-93} + v_{i-92} \cdot v_{i-91} + v_{i-66} + w_{i-78}, \\
z_i &= w_{i-84} + w_{i-83} \cdot w_{i-82} + w_{i-69} + z_{i-87}.
\end{align*}
Here, \( \mathbf{v} \), \( \mathbf{w} \), and \( \mathbf{z} \) are the shift registers of lengths 84, 93, and 111 bits, respectively. After this initialization phase, Trivium produces the keystream vector \( \mathbf{\bar{k}} = (\bar{k}_1, \bar{k}_2, \dots) \) using the following equation:
\begin{equation}
\bar{k}_i = z_{i-111} + v_{i-93} + w_{i-84} + z_{i-66} + v_{i-66}.    
\label{eq:nextBit}
\end{equation}

\input{Figs/conv_figures}

Below is a general description of the functionality provided by an FHE-friendly stream cipher.
\begin{itemize}
    \item \(  (\mathbf{k}, \mathbf{IV}) \stackrel{\$}{\leftarrow} \mathsf{E.KeyGen}(1^{\lambda_{\mathsf{Sym}}}) \). The key vector \( \mathbf{k} \) and initialization vector \( \mathbf{IV} \) are generated based on the security parameter \( \lambda_{\mathsf{Sym}} \).
    
    \item \( \mathsf{E.Init}(\mathbf{k}, \mathbf{IV}) \). The stream cipher is initialized with the key \( \mathbf{k} \) and initialization vector \( \mathbf{IV} \). This step loads the internal state, which typically consists of shift registers, to prepare for keystream generation.
    
    \item \( \mathbf{\bar{k}} \leftarrow  \mathsf{E.KeyStream}(l) \). A keystream vector $ \mathbf{\bar{k}} = (\bar{k}_1,...,$ $ \bar{k}_l )$ is generated by the stream cipher, according to the recurrence relation (Eq.~\ref{eq:nextBit}).
    
    \item \( c_i \leftarrow \mathsf{E}(m_i, \bar{k}_i)  \). Each ciphertext bit \( c_i \) is produced by XORing the message bit \( m_i \) with the corresponding keystream bit \( \bar{k}_i \): \( c_i = m_i \oplus \bar{k}_i \).
\end{itemize}

%% file: Sections/Defs/lwe_sample.tex
\begin{definition}[LWE Sample over $\mathbb{T}$]
An LWE sample over the torus $\mathbb{T} = \mathbb{R} / \mathbb{Z}$ is defined as a pair $(\mathbf{a}, b)$, where $\mathbf{a}$ is a vector sampled uniformly at random from $\mathbb{T}^n$, and $b = \langle \mathbf{a}, \mathbf{s} \rangle + e$. Here, $\mathbf{s}$ is a secret key vector sampled from a key distribution over $\mathbb{B}^n$, where $\mathbb{B} = \{0, 1\}$, and $e$ is an error term drawn from a Gaussian distribution $\chi$ over $\mathbb{R}$.
\end{definition}

%% file: Sections/Defs/decisional.tex
\begin{definition}[Decisional LWE Problem over $\mathbb{T}$]
Given an arbitrary number of samples, the decisional LWE problem over the torus $\mathbb{T}$ is to distinguish between samples of the form $(\mathbf{a}, b)$, where $(\mathbf{a}, b)$ is an LWE sample with $b = \mathbf{a} \cdot \mathbf{s} + e$ for a fixed secret key $\mathbf{s}$, and samples drawn uniformly from $\mathbb{T}^n \times \mathbb{T}$.
\end{definition}

%% file: Sections/Defs/pubkey.tex
\begin{definition}[Public Key of LWE Sample]
The public key set $\mathsf{pk}_{\mathbf{m}}$ of an LWE sample is defined as $\{(\mathbf{a}_i, b_i)\}_{i \in [n_{\mathbf{m}}]}$, where each $\mathbf{a}_i \stackrel{\$}{\leftarrow} \mathbb{T}^n$ is a vector sampled uniformly from $\mathbb{T}^n$, and $b_i = \mathbf{a}_i \cdot \mathbf{s} + e_i$. Here, $\mathbf{s} = (s_1, \dots, s_n)$ is the secret key vector, uniformly drawn from $\mathbb{B}^n$, and $e_i$ is an error term sampled from a Gaussian distribution $\chi$ over $\mathbb{R}$. $n_{\mathbf{m}}$ denotes the number of elements in the vector $\mathbf{m}$.
\end{definition}

%% file: Sections/Defs/rem_1.tex
\begin{remark}
The notation \(\mathsf{Enc}(\mathbf{m}, \mathsf{pk}_{\mathbf{m}})\) denotes a set of \(n_{\mathbf{m}}\) LWE samples in the public key set \(\mathsf{pk}_{\mathbf{m}}\) associated with the message vector \(\mathbf{m} = \{m_1, \dots, m_{n_{\mathbf{m}}}\}\), where each \(m_i \in \mathbb{B}\). For each \(m_i\), the corresponding LWE sample \(\mathsf{pk}_{\mathbf{m}, i} = (\mathbf{a}_i, b_i) \in \mathsf{pk}_{\mathbf{m}}\) is used for encryption, i.e., \(\mathsf{Enc}(m_i, \mathsf{pk}_{\mathbf{m}, i})\).
\end{remark}

%% file: Sections/Defs/rem_2.tex
\begin{remark}
The notation \(\mathsf{Enc}(\mathbf{m})\) omits \(\mathsf{pk}_{\mathbf{m}}\) for simplicity and readability. Thus, \(\mathsf{Enc}(\mathbf{m})\) is understood as \(\mathsf{Enc}(\mathbf{m}, \mathsf{pk}_{\mathbf{m}})\) unless otherwise specified.
\end{remark}

%% file: Figs/conv_figures.tex
% \begin{figure*}[!t]
%     \centering
%     % First subfigure
%     \subfloat[Cancelable biometric system]{%
%         \includegraphics[height=0.25\textheight]{Figs/st-fhe.png}%
%         \label{fig:cancel-bio}}
%     \hfil
%     % Second subfigure
%     \subfloat[FHE-based biometric system]{%
%         \includegraphics[height=0.25\textheight]{Figs/orig-tc.png}%
%         \label{fig:fhe-bio}}
%     % Overall caption
%     \caption{Conventional biometric systems: comparison between cancelable biometrics and traditional FHE-based systems. (a) Cancelable biometric system. (b) FHE-based biometric system.}
%     \label{fig:conven}
% \end{figure*}

% \begin{figure*}[!t]
%     \centering
%     % First subfigure
%     \subfloat[Cancelable biometric system]{%
%         \includegraphics[height=0.27\textheight]{Figs/st-fhe.pdf}%
%         \label{fig:cancel-bio}}
%     \hfil
%     % Second subfigure
%     \subfloat[FHE-based biometric system]{%
%         \includegraphics[height=0.27\textheight]{Figs/orig-tc.pdf}%
%         \label{fig:fhe-bio}}
%     \hfil
%     % third subfigure
%     \subfloat[Ours]{%
%         \includegraphics[height=0.27\textheight]{Figs/ours-btf.pdf}%
%         \label{fig:ours-btf}}
%     % Overall caption
%     \caption{Conventional biometric systems: comparison between cancelable biometrics and traditional FHE-based systems. (a) Cancelable biometric system. (b) FHE-based biometric system.}
%     \label{fig:conven}
% \end{figure*}

\begin{figure*}[t]
    \centering
    \subfloat[\textsf{ST-FHE}]{
        \includegraphics[height=0.25\textheight]{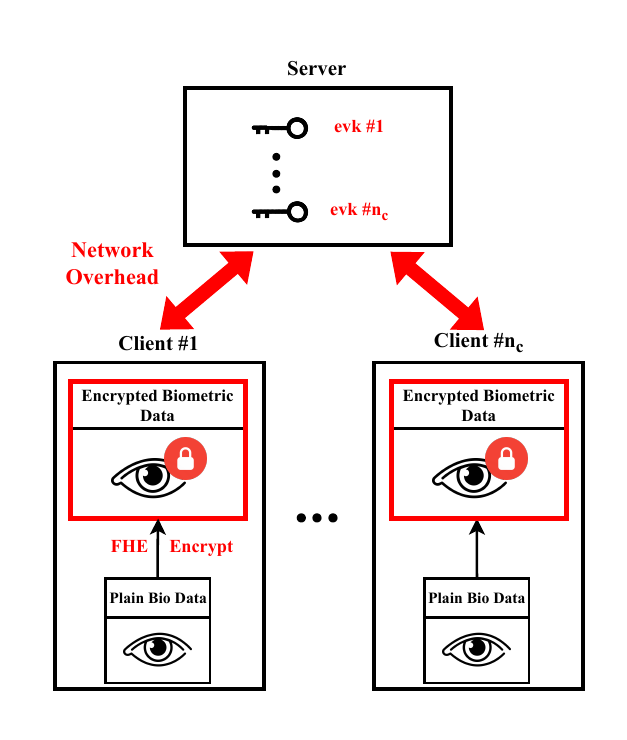}
        \label{fig:st-fhe}}
    \hfil
    \subfloat[\textsf{Orig-TC}]{
        \includegraphics[height=0.25\textheight]{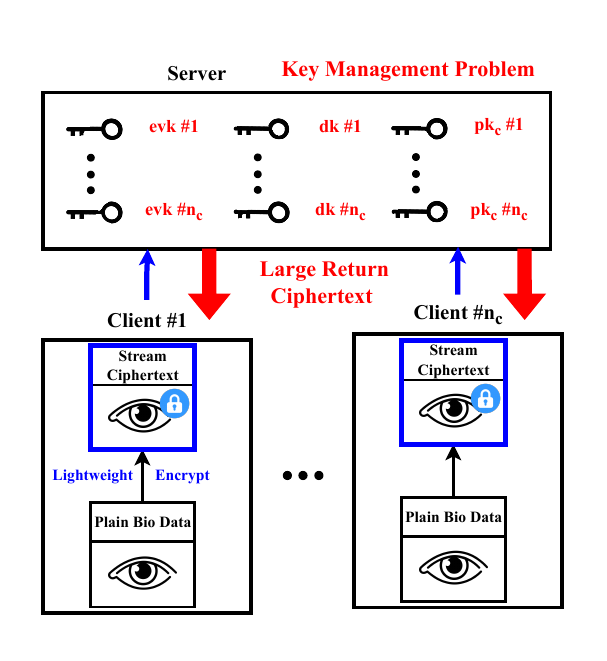}
        \label{fig:orig-tc}}
    \hfil
    \subfloat[\textsf{BTF}]{
        \includegraphics[height=0.25\textheight]{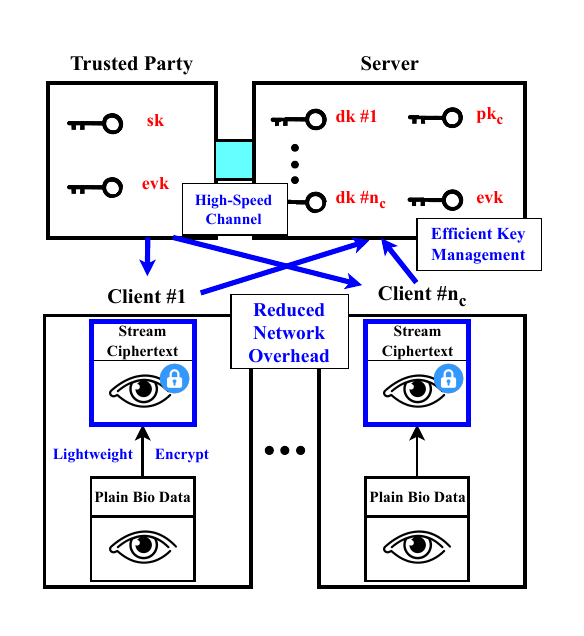}
        \label{fig:ours-btf}}
    
    \caption{Comparison of biometric authentication architectures. We illustrate three designs: (a) standard FHE-based authentication (\textsf{ST-FHE}), (b) original transciphering over FHE (\textsf{Orig-TC}), and (c) our proposed Bidirectional Transciphering Framework (\textsf{BTF}). Our work minimizes network overhead through reduced returning ciphertext size and enables efficient key management via a trusted party and a high-speed channel.}
    \label{fig:conven}
\end{figure*}

%% file: Sections/03_problem.tex
\section{Challenges and Overview of Our Approach}
\label{sec:problem}

\subsection{Problems in Traditional Biometric Authentication Systems}

We classify traditional biometric authentication systems into two representative categories: Standard FHE-based authentication (\textsf{ST-FHE}) and original transciphering over FHE (\textsf{Orig-TC}) (see Fig.~\ref{fig:st-fhe} and Fig.~\ref{fig:orig-tc}). While both approaches aim to enable privacy-preserving biometric matching, they suffer from critical limitations such as high network overhead and complex key management. We summarize and contrast these limitations below.

\noindent\textbf{Standard FHE-based Biometric Authentication (\textsf{ST-FHE}).} \textsf{ST-FHE} systems are conceptually straightforward but impractical for real-world deployment due to severe network and storage overhead. For instance, encrypting a single iris feature vector of size 0.25 KB using LWE-based TFHE produces a ciphertext of approximately 4.93 MB under our implementation (see Section~\ref{sec:eval})---a roughly 20,000$\times$ expansion. This enormous size leads to significant network congestion even during registration or verification for a single user. Furthermore, each client must transmit a large evaluation key \textsf{evk} to the server, which in our setting amounts to 41.6 MB. 

\noindent\textbf{Original Transciphering over FHE (\textsf{Orig-TC}).} 
\textsf{Orig-TC} systems offer a promising approach to mitigating ciphertext transmission overhead by allowing clients to send lightweight stream ciphertexts instead of heavy FHE ciphertexts. However, this comes with nontrivial costs. To enable homomorphic evaluation, the server must re-encrypt the received stream ciphertexts under the client-specific FHE public key $\mathsf{pk}_c$ in order to perform homomorphic decryption and obtain $\mathsf{Enc}(m)$ using the associated decryption key $\mathsf{dk}$. As a result, the server must be provisioned with a set of FHE keys—specifically, $\mathsf{pk}_c, \mathsf{dk}$—which must be transmitted from each client and stored by the server. This per-client key requirement introduces both communication and storage overhead, particularly in large-scale deployments. These issues are further analyzed in Section~\ref{sec:network_comp} and Section~\ref{sec:key}.

\noindent\textbf{Lack of Result Verifiability.} A shared limitation of both \textsf{ST-FHE} and \textsf{Orig-TC} is the reliance on the client to decrypt the final authentication result. Since the result remains encrypted under the client’s secret key, only the client can decrypt it and determine whether authentication was successful. This opens the possibility for malicious clients to falsely claim successful authentication, as the server cannot independently verify the result. This lack of verifiability compromises the integrity of the authentication process and highlights the need for a third party that can securely decrypt and report the result without exposing sensitive biometric data.

\subsection{Brief Explanation of Our Work}
\label{subsec:brief_expl}

Our proposed approach aims to reduce the \emph{network congestion} caused by large ciphertexts and key sizes during both the transmission and return phases of biometric authentication systems (see Fig.~\ref{fig:ours-btf}).

\noindent\textbf{Trusted Party with Double Encryption.} The key architectural change in our design is the introduction of a trusted party. Our novelty lies in ensuring that neither the trusted party nor the server can access the client’s biometric data independently. Specifically, the client performs double encryption of the stream cipher key $\mathbf{k}$: first under the FHE public key $\mathsf{pk}_{\mathbf{k}}$, which is shared with the trusted party, and then under a symmetric key $\mathbf{k'}$, which is shared with the server. Under the non-collusion assumption, this setup guarantees that the biometric data remains inaccessible to any third party alone, preserving the privacy guarantees of FHE while relaxing the trust model for practical deployment.

\noindent\textbf{Efficient Key Management and Scalability.} Beyond double encryption, our framework improves key management. In contrast to traditional approaches where clients generate and transmit unique FHE key pairs, our system allows the trusted party to generate and distribute shared public and evaluation keys. These keys, particularly the large evaluation key $\mathsf{evk}$, can be reused across clients and securely shared with the server. Since communication between the trusted party and the server often occurs over a high-speed channel, the network overhead associated with key transmission is further minimized.

%% file: Sections/04_our_model.tex
\section{Design of BTF Protocol}
\label{sec:our_model}

\noindent \textbf{Assumption.} Our work is designed to minimize network overhead while preserving privacy during biometric authentication. It involves three main parties: the client (\(\mathcal{C}\)), the server (\(\mathcal{S}\)), and the trusted party (\(\mathcal{TP}\)). Both the server and trusted party are assumed to be non-colluding, semi-honest entities; they follow the protocol but are curious about the client’s biometric data.

\noindent \textbf{Protocol Structure.} The \textsf{BTF} framework operates across three main stages: the setup stage, registration stage (\textsf{RS}), and verification stage (\textsf{VS}). The setup stage includes two phases: the Key Distribution Phase (\textsf{KDP}) and the Initialization Phase (\textsf{INP}). In \textsf{KDP}, both FHE and stream cipher keys are distributed. \textsf{INP} manages the initialization of the stream cipher at $\mathcal{C}$ and its FHE counterpart at $\mathcal{S}$. The \textsf{RS} handles client biometric data registration, while the \textsf{VS} performs the authentication procedure (see Alg.~\ref{alg:overview}).

\input{Algs/overview_alg}

\subsection{Setup Stage}
\subsubsection{Key Distribution Phase (KDP)}

% ---- key dist phase
\begin{figure}[!htb] 
  \centering 
  \includegraphics[width=0.45\textwidth]{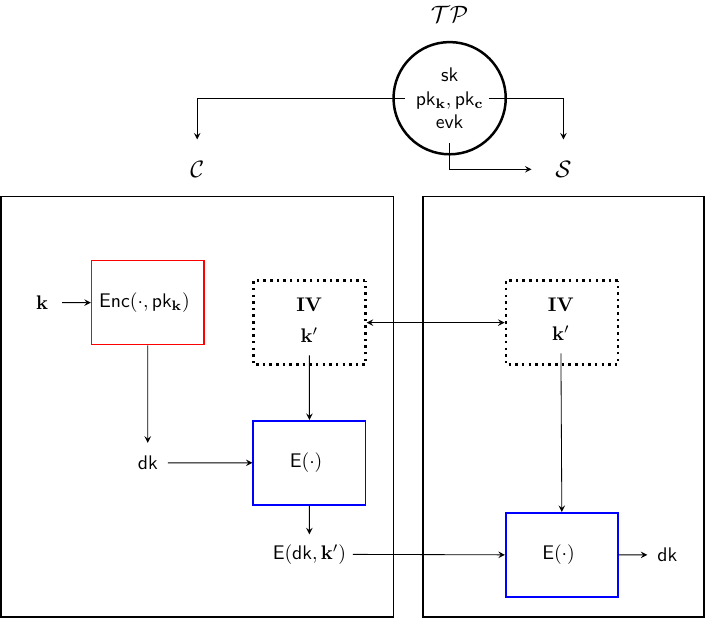}
  \caption{Overview of Key Distribution Phase.} 
  \label{fig:kdp} 
\end{figure}

\textsf{KDP} ensures the generation and distribution of cryptographic keys (see Fig.~\ref{fig:kdp}). The FHE keys are distributed to both $\mathcal{C}$ and $\mathcal{S}$, while the secret key \(\mathsf{sk}\) remains with $\mathcal{TP}$. This phase involves the following steps (see Alg.~\ref{alg:kdp}).

First, $\mathcal{TP}$ generates the FHE key set based on the security parameter $\lambda_{\mathsf{FHE}}$. Then, it securely distributes the public key $\mathsf{pk}_{\mathbf{k}}$ to $\mathcal{C}$ for encrypting the stream cipher key $\mathbf{k}$, and sends both the public key $\mathsf{pk}_{\mathbf{c}}$—used for encrypting stream ciphertext—and the evaluation key $\mathsf{evk}$ to $\mathcal{S}$.

\input{Algs/kdp_alg}

$\mathcal{C}$ selects a stream cipher $\mathsf{E}$ (e.g., Trivium or Kreyvium) based on the required security level $\lambda_{\mathsf{Sym}}$. The client then generates a random symmetric key $\mathbf{k}$ and an initialization vector $\mathbf{IV}$ for this cipher. To secure the symmetric key, $\mathcal{C}$ encrypts $\mathbf{k}$ with the public key $\mathsf{pk}_{\mathbf{k}}$, resulting in the FHE ciphertext of $\mathbf{k}$, denoted as $\mathsf{Enc}(\mathbf{k}, \mathsf{pk}_{\mathbf{k}})$, or $\mathsf{dk}$ for simplicity.

Next, $\mathcal{C}$ generates a secondary symmetric key $\mathbf{k'}$ for double encryption, selected uniformly at random as $\mathbf{k'} \stackrel{\$}{\leftarrow} 1^{l_{\mathsf{dk}}}$, where $l_{\mathsf{dk}}$ denotes the bit length of the homomorphic decryption key $\mathsf{dk}$. The client then encrypts $\mathsf{dk}$ using the stream cipher $\mathsf{E}$ with the key $\mathbf{k'}$, producing the doubly encrypted key $\mathsf{E}(\mathsf{dk}, \mathbf{k'})$. $\mathcal{C}$ shares $\mathbf{k'}$ and the initialization vector $\mathbf{IV}$ with $\mathcal{S}$ over a secure TLS channel, enabling the server to decrypt the stream-encrypted key $\mathsf{E}(\mathsf{dk}, \mathbf{k'})$.

\subsubsection{Initialization Phase (INP)}

% ---- key dist phase
\begin{figure}[!htb] 
  \centering 
  \includegraphics[width=0.45\textwidth]{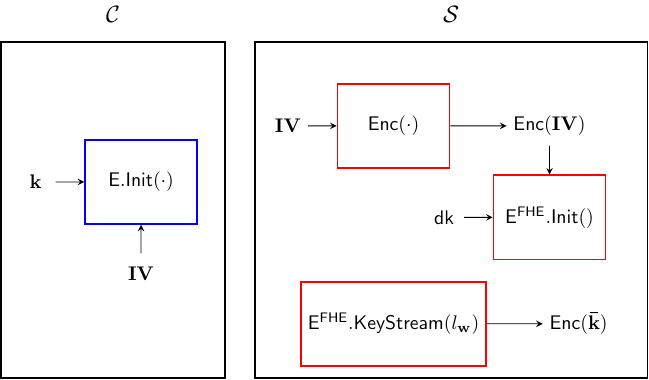}
  \caption{Overview of Initialization Phase.} 
  \label{fig:inp} 
\end{figure}

In \textsf{INP}, $\mathcal{C}$ and $\mathcal{S}$ initialize both the symmetric stream cipher $\mathsf{E}$ and its homomorphic counterpart $\mathsf{E^{FHE}}$ (see Alg.~\ref{alg:inp} and Fig.~\ref{fig:inp}). Specifically, $\mathcal{C}$ initializes $\mathsf{E}$ using the key $\mathbf{k}$ and initialization vector $\mathbf{IV}$. On the server side, $\mathcal{S}$ encrypts $\mathbf{IV}$ under the FHE public key to obtain $\mathsf{Enc}(\mathbf{IV})$, and then uses the homomorphic decryption key $\mathsf{dk}$ to initialize $\mathsf{E^{FHE}}$. The server then generates a sequence of homomorphically encrypted key stream bits, denoted as $\mathsf{Enc}(\bar{\mathbf{k}})$, of length $l_{\mathbf{w}}$, where $l_{\mathbf{w}}$ is the bit-length of the biometric feature vector.

\input{Algs/inp_alg}

\subsection{Registration Stage (RS)}

% ---- registration Stage
\begin{figure}[!htb] 
  \centering 
  \includegraphics[width=0.45\textwidth]{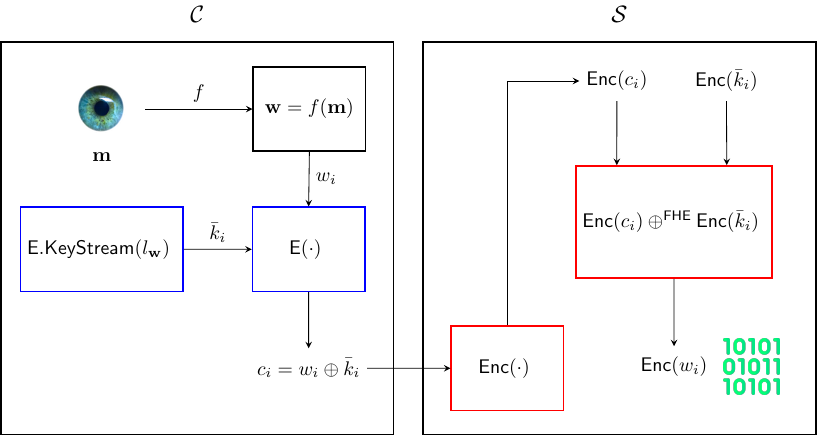}
  \caption{Overview of Registration Stage.} 
  \label{fig:rp} 
\end{figure}

In \textsf{RS}, the client’s biometric data is processed, encrypted, and securely stored at the server (see Fig.~\ref{fig:rp} and Alg.~\ref{alg:rp}). $\mathcal{C}$ begins by using a biometric device to extract raw biometric data such as an iris, face, or fingerprint sample, denoted as $\mathbf{m} \in \mathcal{M}$. The data $\mathbf{m}$ is then transformed via a function $f$ into a feature vector $\mathbf{w} \in \mathcal{W}$, where $f: \mathcal{M} \rightarrow \mathcal{W}$ may involve techniques such as hashing or feature extraction. $\mathcal{C}$ generates a key stream $\bar{\mathbf{k}} \stackrel{\$}{\leftarrow} \mathsf{E.KeyStream}(l_{\mathbf{w}})$, where $l_{\mathbf{w}}$ is the bit-length of the feature vector $\mathbf{w}$. Each bit $w_i$ is then XORed with the corresponding key stream bit $\bar{k}_i$, resulting in the encrypted feature vector $\mathbf{c} = [c_1, c_2, ..., c_{l_{\mathbf{w}}}]$, which is transmitted to $\mathcal{S}$.

\input{Algs/rp_alg}

% Upon receiving the encrypted feature vector, the server encrypts each component $c_i$ using the public key $\mathsf{pk}$ to produce $\mathsf{Enc}(c_i, \mathsf{pk})$. The server also generates homomorphic key streams $\mathsf{Enc}(\mathbf{s}, \mathsf{pk}) \stackrel{\$}{\leftarrow} \mathsf{E^{FHE}.KeyStream}(l_{\mathbf{w}})$. The server then performs homomorphic stream cipher decryption by evaluating $\mathsf{Eval}(\mathsf{E}^{-1}, \mathsf{Enc}(c_i, \mathsf{pk}), \mathsf{Enc}(s_i, \mathsf{pk}), \mathsf{evk})$ using the evaluation key $\mathsf{evk}$, which computes the homomorphic XOR. This operation yields $\mathsf{Enc}(w_i, \mathsf{pk})$, the homomorphic encryption of the feature vector. Finally, the encrypted feature vector $\mathsf{Enc}(\mathbf{w}, \mathsf{pk})$ is securely stored as a biometric template on the server.

Upon receiving the encrypted feature vector $\mathbf{c}$ from $\mathcal{C}$, $\mathcal{S}$ encrypts each element $c_i = w_i \oplus \bar{k}_i$ using the public key $\mathsf{pk}_{\mathbf{c}}$, resulting in ciphertexts $\mathsf{Enc}(c_i)$. (For simplicity, we omit $\mathsf{pk}$ from notation in the remainder of this section.) 

Next, $\mathcal{S}$ generates the corresponding homomorphic key stream $\mathsf{Enc}(\bar{\mathbf{k}}) \stackrel{\$}{\leftarrow} \mathsf{E^{FHE}.KeyStream}(l_{\mathbf{w}})$. $\mathcal{S}$ then performs homomorphic stream cipher decryption by evaluating the inverse of the encryption function: $\mathsf{Enc}(w_i) \leftarrow \mathsf{Eval}(\mathsf{E}^{-1}, \mathsf{Enc}(c_i), \mathsf{Enc}(\bar{k}_i), \mathsf{evk})$. Since this corresponds to a bitwise homomorphic XOR operation, it can be expressed as: $\mathsf{Enc}(w_i) = \mathsf{Enc}(c_i) \oplus^{\mathsf{FHE}} \mathsf{Enc}(\bar{k}_i)$, where $\oplus^{\mathsf{FHE}}$ denotes XOR evaluated over ciphertexts.

This operation yields the homomorphic encryption of the feature vector element $w_i$, since $c_i \oplus \bar{k}_i = (w_i \oplus \bar{k}_i) \oplus \bar{k}_i = w_i$. Finally, the encrypted feature vector $\mathsf{Enc}(\mathbf{w})$ is securely stored as a biometric template at $\mathcal{S}$.

\subsection{Verification Stage (VS)}

% ---- verify stage 
\begin{figure}[!htb] 
  \centering 
  \includegraphics[width=0.45\textwidth]{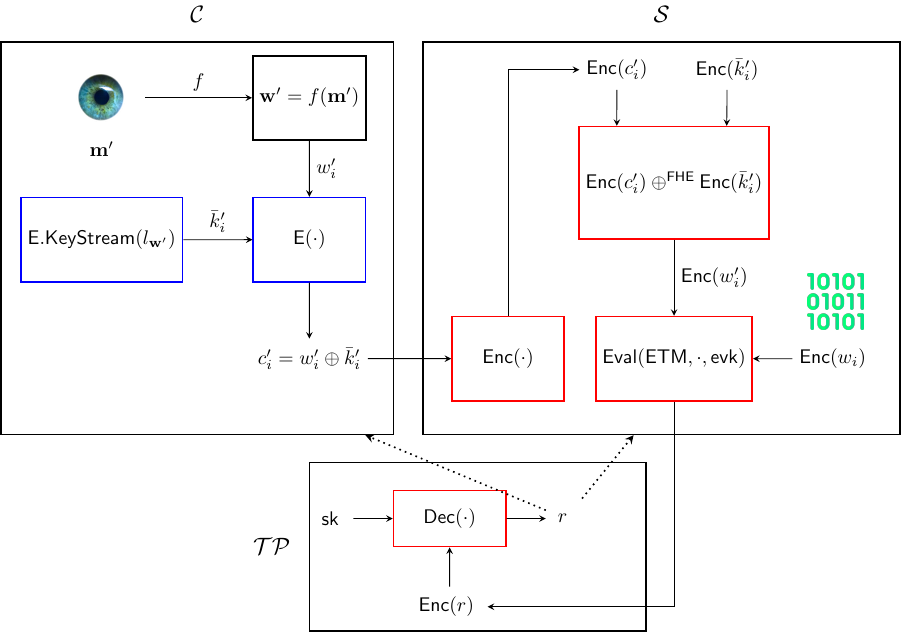}
  \caption{Overview of Verification Stage.} 
  \label{fig:vs} 
\end{figure}

In \textsf{VS}, $\mathcal{C}$ follows a process similar to the registration stage, with the primary difference occurring in the final matching and authentication steps (see Fig.~\ref{fig:vs} and Alg.~\ref{alg:vs}). First, $\mathcal{C}$ captures a new biometric sample $\mathbf{m'} \in \mathcal{M}$, transforms it into a feature vector $\mathbf{w'} = f(\mathbf{m'})$, and generates a corresponding key stream $\bar{\mathbf{k'}} \stackrel{\$}{\leftarrow} \mathsf{E.KeyStream}(l_{\mathbf{w'}})$. The feature vector is then encrypted using the same XOR-based method as in \textsf{RS}, resulting in $\mathbf{c'} = [c'_1, \ldots, c'_{l_{\mathbf{w'}}}]$, which is sent to $\mathcal{S}$.

\input{Algs/vs_alg}

Upon receiving $\mathbf{c'}$, $\mathcal{S}$ encrypts each element under the FHE public key $\mathsf{pk}_{\mathbf{c'}}$ and generates the corresponding homomorphic key stream $\mathsf{Enc}(\bar{\mathbf{k'}}) \stackrel{\$}{\leftarrow} \mathsf{E^{FHE}.KeyStream}(l_{\mathbf{w'}})$. Homomorphic stream cipher decryption is then performed as in \textsf{RS} to recover the encrypted feature vector $\mathsf{Enc}(\mathbf{w'})$.

Next, $\mathcal{S}$ evaluates the error tolerance matching algorithm $\mathsf{Eval}(\mathsf{ETM}, \mathsf{Enc}(\mathbf{w}), \mathsf{Enc}(\mathbf{w'}), \mathsf{evk})$ to determine the authentication result. The matching algorithm can be instantiated with various distance metrics such as Euclidean distance~\cite{bio-tfhe} or Hamming distance~\cite{bio-partial}. The result is an FHE-encrypted bit $\mathsf{Enc}(r)$, where $r \in \{0, 1\}$ indicates whether the authentication was successful.

Finally, $\mathcal{S}$ sends $\mathsf{Enc}(r)$ to the trusted party $\mathcal{TP}$, which decrypts it using the secret key $\mathsf{sk}$ and communicates the result $r$ to both $\mathcal{C}$ and $\mathcal{S}$, thereby concluding the verification phase.

%% file: Algs/overview_alg.tex
\begin{algorithm}[!ht]
\label{alg:btf}
\DontPrintSemicolon
  
  \KwInput{Biometric sample $\mathbf{m}$ or $\mathbf{m'}$, $\lambda_{\mathsf{FHE}}, \lambda_{\mathsf{Sym}}$}
  \KwOutput{Authentication result to $\mathcal{C}$ and $\mathcal{S}$}
  
  \SetUp{
    \tcc{Key Distribution Phase (KDP)}
    $\mathsf{BTF.KeyDist}(\lambda_{\mathsf{FHE}}, \lambda_{\mathsf{Sym}})$ \;
    \tcc{Initialization Phase (INP)}
    $\mathsf{BTF.Init}(\mathbf{k}, \mathbf{IV}, \mathsf{dk}, \mathsf{evk})$\;
  }
  \RegisterStage{
    $\mathsf{BTF.Register}(\mathbf{m}, \mathsf{pk}_{\mathbf{c}}, \mathsf{evk})$ \tcp{Register template}
  }  
  \VerificationStage{
    \tcc{Notify auth. result to $\mathcal{C}$ and $\mathcal{S}$}
    $\mathsf{BTF.Verify}(\mathbf{m'}, \mathsf{pk}_{\mathbf{c'}}, \mathsf{evk})$\;
  }
\caption{$\mathsf{BTF}$ Overview}
\label{alg:overview}
\end{algorithm}

%% file: Algs/kdp_alg.tex
\begin{algorithm}[!ht]
\DontPrintSemicolon
  
  \KwInput{Security parameters $\lambda_{\mathsf{FHE}}$, $\lambda_{\mathsf{Sym}}$}
  \KwOutput{Keys distributed to $\mathcal{C}$ and $\mathcal{S}$}

  \TrustedParty{
    % \tcc{Generate FHE keys and distribute to $\mathcal{C}$ and $\mathcal{S}$}
    $\mathsf{sk}, \mathsf{pk}_{\mathbf{k}}, \mathsf{pk}_{\mathbf{c}}, \mathsf{evk} \stackrel{\$}{\leftarrow} \mathsf{Enc.KeyGen(1^{\lambda_{FHE}})}$\;
    Distribute $\mathsf{pk}_{\mathbf{k}}$ to $\mathcal{C}$, and $\mathsf{pk}_{\mathbf{c}}$, $\mathsf{evk}$ to $\mathcal{S}$\;
  }

  \Client{
    Select stream cipher $\mathsf{E}$ based on $\lambda_{\mathsf{Sym}}$\;
    \( \mathbf{k}, \mathbf{IV} \stackrel{\$}{\leftarrow} \mathsf{E.KeyGen}(1^{\lambda_{\mathsf{Sym}}}) \)\;

    \tcc{$\mathsf{dk}$: homomorphic $\mathsf{SC}$ decryption key}
    $\mathsf{dk} \leftarrow \mathsf{Enc}(\mathbf{k}, \mathsf{pk}_{\mathbf{k}})$\;
    
     $\mathbf{k'} \stackrel{\$}{\leftarrow} 1^{l_{\mathsf{dk}}}$\;
     
     \tcc{Double encryption of $\mathsf{dk}$}
    Encrypt $\mathsf{dk}$ with $\mathbf{k'}$ to obtain $\mathsf{E}(\mathsf{dk}, \mathbf{k'})$\;
    Send $\mathsf{E}(\mathsf{dk}, \mathbf{k'})$ to $\mathcal{S}$\;
    
    Send $\mathbf{IV}, \mathbf{k'}$ to $\mathcal{S}$ via TLS\;
  }

  \Server{
    % \tcc{Decrypts homomorphic decryption key}
    Decrypt $\mathsf{E}(\mathsf{dk}, \mathbf{k'})$ using $\mathbf{k'}$ to obtain $\mathsf{dk}$\;
  }

\caption{$\mathsf{BTF.KeyDist}(\lambda_{\mathsf{FHE}}$, $\lambda_{\mathsf{Sym}})$}
\label{alg:kdp}
\end{algorithm}

%% file: Algs/inp_alg.tex
\begin{algorithm}[!ht]
\DontPrintSemicolon
  
  \KwInput{$\mathbf{k}$, $\mathbf{IV}$, $\mathsf{dk}$, $\mathsf{evk}$}
  \KwOutput{Initialized (plain and FHE) stream ciphers}

  \Client{
    % \tcc{Initializes symmetric cipher $\mathsf{E}$}
    $\mathsf{E.Init}(\mathbf{k}, \mathbf{IV})$
  }

  \Server{
    \(\mathsf{ct}_{\mathbf{IV}} \leftarrow \mathsf{Enc}(\mathbf{IV})\)\;

    \tcc{Initialize homomorphic stream cipher}
    $\mathsf{E^{FHE}.Init}(\mathsf{ct}_{\mathbf{IV}}, \mathsf{dk}, \mathsf{evk})$ \;

    $\mathsf{Enc(\mathbf{\bar{k}})} \leftarrow \mathsf{E^{FHE}.KeyStream}(l_{\mathbf{w}})$\;
  }

\caption{$\mathsf{BTF.Init}(\mathbf{k}, \mathbf{IV}, \mathsf{dk}, \mathsf{evk})$}
\label{alg:inp}
\end{algorithm}

%% file: Algs/rp_alg.tex
\begin{algorithm}[!ht]
\DontPrintSemicolon
  
  \KwInput{Biometric data $\mathbf{m}$, $\mathsf{pk}_{\mathbf{c}}, \mathsf{evk}$}
  \KwOutput{Encrypted feature vector $\mathsf{Enc}(\mathbf{w})$}
  
  \Client{
    % \tcc{Biometric data extraction and feature transformation}
    $\mathbf{m}$ is extracted using a biometric device\;
    \tcc{Transform $\mathbf{m}$ into feature vector}
    $\mathbf{w} \leftarrow f(\mathbf{m})$ 
    
    % \tcc{Key stream generation and feature vector encryption}
    $\mathbf{\bar{k}} \stackrel{\$}{\leftarrow} \mathsf{E.KeyStream}(l_{\mathbf{w}})$ \tcp{$l_{\mathbf{w}}:$ length of $\mathbf{w}$}
    \tcc{Encrypt each bit of feature $\mathbf{w}$}
    % $c_i = w_i \oplus s_i$ for $i = 1$ to $l_{\mathbf{w}}$\;
    $\mathbf{c} \leftarrow \mathsf{E}(\mathbf{w}, \mathbf{\bar{k}})$\;
    Send $\mathbf{c} = [c_1, c_2, ..., c_{l_{\mathbf{w}}}]$ to $\mathcal{S}$\;
  }

    \Server{
      $\mathsf{Enc}(\mathbf{c}, \mathsf{pk}_{\mathbf{c}})$\;
    
      \tcc{Homomorphic $\mathsf{SC}$ decryption ($\mathsf{evk}$ req.)}
      \For{$i = 1$ \KwTo $l_{\mathbf{w}}$}{
        $\mathsf{Enc}(w_i) \leftarrow \mathsf{Enc}(c_i) \oplus^{\mathsf{FHE}} \mathsf{Enc}(\bar{k}_i)$\;
      }
    
      Store template $\mathsf{Enc}(\mathbf{w})$ at $\mathcal{S}$\;
    }

\caption{$\mathsf{BTF.Register}(\mathbf{m}, \mathsf{pk}_{\mathbf{c}}, \mathsf{evk})$}
\label{alg:rp}
\end{algorithm}

%% file: Algs/vs_alg.tex
\begin{algorithm}[!ht]
\DontPrintSemicolon
  
  \KwInput{Biometric data $\mathbf{m'}$, $\mathsf{pk}_\mathbf{c'}$, $\mathsf{evk}$}
  \KwOutput{Authentication result to $\mathcal{C}$ and $\mathcal{S}$}
  
  \Client{
    Capture new biometric data $\mathbf{m'}$\;
    $\mathbf{w'} = f(\mathbf{m'})$\;
    $\mathbf{\bar{k'}} \stackrel{\$}{\leftarrow} \mathsf{E.KeyStream}(l_{\mathbf{w'}})$\;
    \tcc{Encrypt each bit of feature $\mathbf{w'}$}
    \For{$i = 1$ \KwTo $l_{\mathbf{w'}}$}{$c'_i = w'_i \oplus \bar{k'}_i$\;} 
    Send $\mathbf{c'} = [c'_1,..., c_{l_{\mathbf{w'}}}]$ to $\mathcal{S}$\;
  }

  \Server{
    $\mathsf{Enc}(\mathbf{c'}, \mathsf{pk}_\mathbf{c'})$\;
    % \tcc{Generate FHE key streams} 
    $\mathsf{Enc(\mathbf{\bar{k'}})} \stackrel{\$}{\leftarrow} \mathsf{E^{FHE}.KeyStream}(l_{\mathbf{w'}})$\;
    \For{$i = 1$ \KwTo $l_{\mathbf{w'}}$}{
    $\mathsf{Enc}(w'_i) \leftarrow \mathsf{Enc}(c_i) \oplus^{\mathsf{FHE}} \mathsf{Enc}(\bar{k}'_i)$\;}
    \tcc{Error tolerance matching (ETM)}
    $\mathsf{Enc}(r) \leftarrow \mathsf{Eval}(\mathsf{ETM}, \mathsf{Enc}(\mathbf{w}), \mathsf{Enc}(\mathbf{w'}), \mathsf{evk})$\;
    Send $\mathsf{Enc}(r)$ to $\mathcal{TP}$\;
  }

  \TrustedParty{
   $r \leftarrow \mathsf{Dec}(\mathsf{Enc}(r), \mathsf{sk})$\;
    Notify auth. result to both $\mathcal{C}$ and $\mathcal{S}$\;
  }

\caption{$\mathsf{BTF.Verify}(\mathbf{m'}, \mathsf{pk}_\mathbf{c'}, \mathsf{evk})$}
\label{alg:vs}
\end{algorithm}

%% file: Sections/05_network_efficiency.tex
\section{Comparative Network Efficiency}
\label{sec:network_comp}

\input{Tabs/tab_data_move}

In this section, we compare the network efficiency of three biometric authentication models: Standard FHE (\textsf{ST-FHE}), original transciphering over FHE (\textsf{Orig-TC}), and our work (\textsf{BTF}). We begin by outlining the data and key movement patterns in the baseline models, then highlight the communication advantages of our design. We also discuss additional components introduced in our framework, which incur only minor overhead and are outweighed by the gains in scalability and efficiency. Tab.~\ref{tab:key_data_move} provides a comprehensive overview of data and key movement across the three frameworks, including the communication direction (e.g., $\mathcal{C} \rightarrow \mathcal{S}$), the cryptographic scheme of each component (FHE vs. stream cipher), and whether data is stored or used temporarily during processing (marked with $\blacktriangle$).

\subsection{Baseline Communication Workflows}

\noindent \textbf{Standard FHE.} In the $\mathsf{ST-FHE}$ model (see Fig.~\ref{fig:st-fhe}), the client generates the FHE key set $(\mathsf{sk}, \mathsf{pk}_{\mathbf{c}}, \mathsf{evk})$ and transmits the evaluation key $\mathsf{evk}$ to the server during the setup stage. During the registration and verification stages, the client sends the encrypted biometric feature $\mathsf{Enc}(\mathbf{w})$ or $\mathsf{Enc}(\mathbf{w'})$, respectively, to the server. In the verification stage, the server performs homomorphic evaluation and returns the encrypted authentication result $\mathsf{Enc}(r)$ to the client, who then decrypts it using the secret key $\mathsf{sk}$ to obtain the result $r$.

\noindent \textbf{Original Transciphering.} In the $\mathsf{Orig-TC}$ model (see Fig.~\ref{fig:transc-general}), the client generates the FHE key set $(\mathsf{sk}, \mathsf{pk}_{\mathbf{k}}, \mathsf{pk}_{\mathbf{c}}, \mathsf{evk})$, along with a stream cipher key $\mathbf{k}$ and an initialization vector $\mathbf{IV}$. The stream cipher key $\mathbf{k}$ is then encrypted using the FHE public key $\mathsf{pk}_{\mathbf{k}}$, producing the homomorphic decryption key $\mathsf{dk} = \mathsf{Enc}(\mathbf{k}, \mathsf{pk}_{\mathbf{k}})$. During the key distribution phase (\textsf{KDP}), the client transmits $\mathsf{pk}_{\mathbf{c}}$, $\mathsf{evk}$, $\mathsf{dk}$, and $\mathbf{IV}$ to the server.

During the registration and verification stages, the client sends a stream-encrypted biometric feature vector $\mathbf{c} = \mathsf{E}(\mathbf{w}, \bar{\mathbf{k}})$ during registration (or $\mathbf{c'}$ during verification) to the server. The server then performs homomorphic stream cipher decryption to recover $\mathsf{Enc}(\mathbf{w})$ or $\mathsf{Enc}(\mathbf{w'})$, respectively. As in the $\mathsf{ST\text{-}FHE}$ model, the server returns the encrypted authentication result $\mathsf{Enc}(r)$ to the client, who decrypts it using the secret key $\mathsf{sk}$ to obtain the final result.

\subsection{Network Efficiency Gains in Our Model}

\noindent \textbf{Efficient Channel from $\mathcal{TP}$ to $\mathcal{S}$.} As described in Section~\ref{subsec:brief_expl}, our model assumes a high-bandwidth, low-latency channel between $\mathcal{TP}$ and $\mathcal{S}$—a reasonable deployment scenario when both parties operate within a shared or co-located infrastructure, such as a cloud-based authentication backend. In contrast, the $\mathcal{C}$-to-$\mathcal{S}$ and $\mathcal{TP}$-to-$\mathcal{C}$ channels typically traverse standard internet connections with bandwidth around 10 Mbps. By comparison, the dedicated channel between $\mathcal{TP}$ and $\mathcal{S}$ can reach 10–100 Gbps using modern data center networking, enabling over 1,000$\times$ faster data transfer. Our design leverages this architectural asymmetry to offload large cryptographic artifacts (e.g., $\mathsf{evk}$, $\mathsf{Enc}(r)$) away from the bandwidth-constrained client path.

\noindent \textbf{Gain \#1: Minimizing $\mathcal{C}$ to $\mathcal{S}$ Key Transfer Overhead.} In both the $\mathsf{ST\text{-}FHE}$ and $\mathsf{Orig\text{-}TC}$ frameworks, a major source of network overhead arises from transmitting the evaluation key $\mathsf{evk}$ and the public key $\mathsf{pk}_{\mathbf{c}}$—used for encrypting the stream ciphertext—over the client-to-server channel. These keys are substantially larger than most other cryptographic data, as we quantify in Section~\ref{sec:eval}. By contrast, our model offloads this burden by having $\mathcal{TP}$ generate and transmit both $\mathsf{evk}$ and $\mathsf{pk}_{\mathbf{c}}$ to $\mathcal{S}$ over the high-bandwidth $\mathcal{TP}$-to-$\mathcal{S}$ channel. Moreover, while Tab.~\ref{tab:key_data_move} presents communication for a single client, the scalability challenge becomes more severe as the number of clients increases—leading to cumulative congestion over the $\mathcal{C}$-to-$\mathcal{S}$ channel due to repeated transmission of heavy FHE keys and encrypted feature vectors. This scalability issue is analyzed in further detail in Section~\ref{sec:key}.

\noindent\textbf{Gain \#2: Improving Network Efficiency for FHE-encrypted Biometric Templates.} In the $\mathsf{ST\text{-}FHE}$ framework, the client transmits the biometric feature vector to the server during both the registration and verification stages. Under the $\mathsf{TFHE128}$ parameter set, encrypting a 0.5 KB biometric feature vector results in a FHE ciphertext of approximately 5.17 MB. Given that verification occurs frequently in biometric authentication, repeatedly transmitting such large FHE-encrypted data imposes substantial network overhead—especially as the number of clients grows. In contrast, our model and the $\mathsf{Orig\text{-}TC}$ framework incur no such overhead, as the biometric feature is transmitted using a symmetric stream cipher without ciphertext expansion.

\noindent \textbf{Gain \#3: Eliminating Overhead from FHE-encrypted Authentication Results.} In both the $\mathsf{ST\text{-}FHE}$ and $\mathsf{Orig\text{-}TC}$ models, the server returns the authentication result to the client as an FHE-encrypted ciphertext $\mathsf{Enc}(r)$, which is approximately 20{,}000$\times$ larger than a single plaintext bit. While the $\mathsf{Orig\text{-}TC}$ framework reduces the size of outbound biometric data by using stream ciphers, it still overlooks the substantial overhead associated with returning ciphertexts. As the number of clients grows, this cumulative server-to-client return-channel load becomes increasingly significant and contributes to network congestion.

By contrast, our bidirectional model offloads FHE decryption to $\mathcal{TP}$, leveraging the efficient $\mathcal{TP}$-to-$\mathcal{S}$ channel to carry the authentication result. The server sends $\mathsf{Enc}(r)$ to $\mathcal{TP}$, which decrypts it using the FHE secret key and returns the plaintext result $r$ to both the client and the server, thereby eliminating the need to transmit large ciphertexts—such as $\mathsf{Enc}(r)$ in the $\mathsf{ST\text{-}FHE}$ and $\mathsf{Orig\text{-}TC}$ models—back to the client.

\subsection{Minor Overhead from Supporting Keys}
\label{subsec:minor_overhead}

\noindent\textbf{Minor Overhead \#1: Additional Public Key $\mathsf{pk}_{\mathbf{k}}$ via $\mathcal{TP}$-to-$\mathcal{C}$ Channel.} 
While $\mathsf{pk}_{\mathbf{k}}$ is locally generated and used by the client in the $\mathsf{Orig\text{-}TC}$ model, it is not transmitted across the network. In contrast, our model requires the trusted party to generate $\mathsf{pk}_{\mathbf{k}}$ and send it to the client during the setup phase. This key is used by the client to encrypt the stream cipher key $\mathbf{k}$, producing the homomorphic decryption key $\mathsf{dk}$ that is forwarded to the server. Although this introduces a one-time communication overhead over the standard $\mathcal{TP}$-to-$\mathcal{C}$ channel, the key $\mathsf{pk}_{\mathbf{k}}$ is incomparable in size to heavier elements such as the evaluation key $\mathsf{evk}$ (see Section~\ref{sec:eval}), is used only temporarily, and does not contribute to recurring network costs.

\noindent \textbf{Minor Overhead \#2: Network Overhead Associated with $\mathsf{dk}$ and $\mathbf{k'}$.} 
Unlike the $\mathsf{ST\text{-}FHE}$ model, both our model and the $\mathsf{Orig\text{-}TC}$ model require the transmission of $\mathsf{dk}$ for homomorphic stream cipher decryption of the encrypted feature vector $\mathsf{Enc}(\mathbf{c})$ to recover $\mathsf{Enc}(\mathbf{w})$. However, the size of $\mathsf{dk}$ is negligible compared to the full FHE-encrypted biometric template transmitted in $\mathsf{ST\text{-}FHE}$.

Our model differs from both $\mathsf{ST\text{-}FHE}$ and $\mathsf{Orig\text{-}TC}$ by introducing an additional encryption layer on the stream cipher key $\mathbf{k}$. After $\mathbf{k}$ is encrypted under the FHE scheme to produce $\mathsf{dk}$, it is further encrypted using a symmetric stream cipher with a session key $\mathbf{k'}$, resulting in $\mathsf{E}(\mathsf{dk}, \mathbf{k'})$. While this extra encryption step adds a small overhead, the size of $\mathsf{E}(\mathsf{dk}, \mathbf{k'})$ is identical to $\mathsf{dk}$ due to the non-expanding nature of symmetric encryption. As a result, the only additional cost is the transmission of the session key $\mathbf{k'}$, which is small in size and sent only once during setup through  $\mathcal{C}$-to-$\mathcal{S}$ channel.

%% file: Tabs/tab_data_move.tex
\begin{table*}[htb!]
\caption{Key and data movement across three FHE schemes for biometric authentication: Standard FHE ($\mathsf{ST-FHE}$), original transciphering over FHE ($\mathsf{Orig-TC}$), and our scheme ($\mathsf{BTF}$). The black triangle ($\blacktriangle$) denotes data temporarily used during processing and not stored locally. Red boxes indicate FHE components, while blue boxes represent symmetric stream cipher components.}
\begin{center}
\scalebox{0.92}{
\begin{tabular}{c c ccccccccccccc}
\toprule
\textbf{Framework} & \textbf{Data} & $\mathsf{sk}$ & $\mathsf{pk}_{\mathbf{k}}$ & $\mathsf{pk}_{\mathbf{c}}$ & $\mathsf{evk}$ & $\mathbf{k}$ & $\mathbf{k'}$ & $\mathsf{dk}$ & $\mathsf{E}(\mathsf{dk}, \mathbf{k'})$ & $\mathbf{c}$ & $\mathsf{Enc}(\mathbf{w})$ & $\mathsf{Enc}(\mathbf{w'})$ & $\mathsf{Enc}(r)$ & $r$ \\
\midrule
\multirow{4}{*}{$\mathsf{ST-FHE}$} 
& $\mathcal{C} \rightarrow \mathcal{S}$ & \xmark & \xmark & \xmark & \fcolorbox{red}{white}{\cmark} & \xmark & \xmark & \xmark & \xmark & \xmark & \fcolorbox{red}{white}{\cmark} & \fcolorbox{red}{white}{\cmark} & \xmark & \xmark \\
& $\mathcal{S} \rightarrow \mathcal{C}$ & \xmark & \xmark & \xmark & \xmark & \xmark & \xmark & \xmark & \xmark & \xmark & \xmark & \xmark & \fcolorbox{red}{white}{\cmark} & \xmark \\
& $\mathcal{C}$ & \fcolorbox{red}{white}{\cmark} & \xmark & $\blacktriangle$ & $\blacktriangle$ & \xmark & \xmark & \xmark & \xmark & \xmark & $\blacktriangle$ & $\blacktriangle$ & $\blacktriangle$ & \xmark \\
& $\mathcal{S}$ & \xmark & \xmark & \xmark & \fcolorbox{red}{white}{\cmark} & \xmark & \xmark & \xmark & \xmark & \xmark & \fcolorbox{red}{white}{\cmark} & $\blacktriangle$ & $\blacktriangle$ & \xmark \\
\bottomrule

\multirow{4}{*}{$\mathsf{Orig-TC}$} & $\mathcal{C} \rightarrow \mathcal{S}$ & \xmark & \xmark & \fcolorbox{red}{white}{\cmark} & \fcolorbox{red}{white}{\cmark} & \xmark & \xmark & \fcolorbox{red}{white}{\cmark} & \xmark & \fcolorbox{blue}{white}{\cmark} & \xmark & \xmark & \xmark & \xmark \\
 & $\mathcal{S} \rightarrow \mathcal{C}$ & \xmark & \xmark & \xmark & \xmark & \xmark & \xmark & \xmark & \xmark & \xmark & \xmark & \xmark & \fcolorbox{red}{white}{\cmark} & \xmark\\
 & $\mathcal{C}$ & \fcolorbox{red}{white}{\cmark} & $\blacktriangle$ & $\blacktriangle$ & $\blacktriangle$ & \fcolorbox{blue}{white}{\cmark} & \xmark & $\blacktriangle$ & \xmark & $\blacktriangle$ & \xmark & \xmark & $\blacktriangle$ & \xmark\\
 & $\mathcal{S}$ & \xmark & \xmark & \fcolorbox{red}{white}{\cmark} & \fcolorbox{red}{white}{\cmark} & \xmark & \xmark & \fcolorbox{red}{white}{\cmark} & \xmark & $\blacktriangle$ & \fcolorbox{red}{white}{\cmark} & $\blacktriangle$ & $\blacktriangle$ & \xmark\\
\bottomrule

\multirow{10}{*}{$\mathsf{Ours \hspace{0.5em}(BTF)}$} & $\mathcal{TP} \rightarrow \mathcal{C}$ & \xmark & \fcolorbox{red}{white}{\cmark} & \xmark & \xmark & \xmark & \xmark & \xmark & \xmark & \xmark & \xmark & \xmark & \xmark & \fcolorbox{blue}{white}{\cmark}  \\
 & $\mathcal{TP} \rightarrow \mathcal{S}$ & \xmark & \xmark & \fcolorbox{red}{white}{\cmark} & \fcolorbox{red}{white}{\cmark} & \xmark & \xmark & \xmark & \xmark & \xmark & \xmark & \xmark & \xmark & \fcolorbox{blue}{white}{\cmark}  \\
 & $\mathcal{C} \rightarrow \mathcal{S}$ & \xmark & \xmark & \xmark & \xmark & \xmark & \fcolorbox{blue}{white}{\cmark} & \xmark & \fcolorbox{blue}{white}{\fcolorbox{red}{white}{\cmark}} & \fcolorbox{blue}{white}{\cmark} & \xmark & \xmark & \xmark & \xmark\\
 & $\mathcal{S} \rightarrow \mathcal{C}$ & \xmark & \xmark & \xmark & \xmark & \xmark & \xmark & \xmark & \xmark & \xmark & \xmark & \xmark & \xmark & \xmark\\
 & $\mathcal{S} \rightarrow \mathcal{TP}$ & \xmark & \xmark & \xmark & \xmark & \xmark & \xmark & \xmark & \xmark & \xmark & \xmark & \xmark & \fcolorbox{red}{white}{\cmark} & \xmark\\
 & $\mathcal{TP}$ & \fcolorbox{red}{white}{\cmark} & $\blacktriangle$ & $\blacktriangle$ & \fcolorbox{red}{white}{\cmark} & \xmark & \xmark & \xmark & \xmark & \xmark & \xmark & \xmark & \xmark & \xmark  \\
 & $\mathcal{C}$ & \xmark & $\blacktriangle$ & \xmark & \xmark & \fcolorbox{blue}{white}{\cmark} & $\blacktriangle$ & $\blacktriangle$ & $\blacktriangle$ & $\blacktriangle$ & \xmark & \xmark & \xmark & $\blacktriangle$  \\
 & $\mathcal{S}$ & \xmark & \xmark & \fcolorbox{red}{white}{\cmark} & \fcolorbox{red}{white}{\cmark} & \xmark & \xmark & \fcolorbox{red}{white}{\cmark} & $\blacktriangle$ & $\blacktriangle$ & \fcolorbox{red}{white}{\cmark} & $\blacktriangle$ & \xmark & $\blacktriangle$  \\
 \bottomrule

\end{tabular}
}
\end{center}
\label{tab:key_data_move}
\end{table*}

%% file: Sections/06_key_management.tex
\section{Key Management and Scalability}
\label{sec:key}

In this section, we compare the three biometric authentication frameworks—$\mathsf{ST\text{-}FHE}$, $\mathsf{Orig\text{-}TC}$, and our work—in terms of key management and system scalability as the number of clients $n_c$ increases. While the previous section focused on network efficiency in a single-client setting, we now analyze how each framework handles the overhead and complexity of managing cryptographic keys and biometric data across many clients (see Tab.~\ref{tab:key_by_party}).

\input{Tabs/key_manage}

\subsection{Standard FHE.} 
\noindent \textbf{Key Management.} In the $\mathsf{ST\text{-}FHE}$ framework, each of the $n_c$ clients independently generates and holds a unique FHE key pair $(\mathsf{pk}_i, \mathsf{sk}_i)$. The server, on the other hand, must store $n_c$ evaluation keys $\{\mathsf{evk}_i\}_{i \in [n_c]}$, one for each client, in order to perform homomorphic operations. Since the client encrypts its own biometric feature vector $\mathbf{w}_i$ locally, it does not need to transmit its public key $\mathsf{pk}_i$ to the server. As a result, $\mathsf{pk}_i$ can be generated and used temporarily by the client without contributing to persistent key management overhead. 

\noindent \textbf{Network Scalability.} The dominant scalability overhead in the $\mathsf{ST\text{-}FHE}$ framework arises not from key management, but from the transmission and storage of FHE-encrypted biometric templates across $n_c$ clients. In addition, the returning authentication results—though smaller—are still FHE ciphertexts and contribute to the overall communication load.

\subsection{Original Transciphering.} 
\noindent \textbf{Key Management.} In the $\mathsf{Orig\text{-}TC}$ model, each client holds a secret key $\mathsf{sk}_i$ for homomorphic encryption and decryption, as well as a symmetric cipher key $\mathbf{k}_i$ for stream cipher encryption. On the server side, key management becomes significantly more complex, as the server must store the evaluation keys $\{\mathsf{evk}_i\}_{i \in [n_c]}$, the homomorphic decryption keys $\{\mathsf{dk}_i\}_{i \in [n_c]}$, and the public keys $\{\mathsf{pk}_{\mathbf{c}_i}\}_{i \in [n_c]}$ used to encrypt the stream ciphertexts.

\noindent \textbf{Network Scalability.} Although $\mathsf{Orig\text{-}TC}$ alleviates the transmission and storage burden of FHE-encrypted biometric templates by using compact stream cipher ciphertexts, it still returns the authentication result as an FHE-encrypted ciphertext via the $\mathcal{S}$-to-$\mathcal{C}$ channel. As the number of clients increases, this returning ciphertext overhead accumulates and contributes to scalability challenges.

\subsection{Our Work.}
\noindent \textbf{Key Management.} In our proposed model, key management is optimized by offloading FHE key generation and distribution to \(\mathcal{TP}\). Each client is only responsible for managing a lightweight symmetric key \(\mathbf{k}_i\), used for encrypting biometric data via a stream cipher. Unlike the $\mathsf{ST\text{-}FHE}$ and $\mathsf{Orig\text{-}TC}$ frameworks, our model does not require each client to generate a unique FHE key pair. Instead, \(\mathcal{TP}\) generates a single global FHE key set \((\mathsf{pk}_{\mathbf{c}}, \mathsf{evk}, \mathsf{sk})\), which is reused across the system. The server synchronizes with \(\mathcal{TP}\) using this key set and maintains only the public key \(\mathsf{pk}_{\mathbf{c}}\), evaluation key \(\mathsf{evk}\), and the client-specific homomorphic decryption keys \(\{\mathsf{dk}_i\}_{i \in [n_c]}\). Meanwhile, \(\mathcal{TP}\) stores the secret key \(\mathsf{sk}\) and the same evaluation key \(\mathsf{evk}\) centrally. 

This design eliminates the need to manage \(\mathcal{O}(n_c)\) public and evaluation key pairs, significantly reducing key management overhead. Since \(\mathsf{evk}\) and \(\mathsf{pk}_{\mathbf{c}}\) are by far the largest components in terms of size, centralizing their management yields substantial scalability gains as the system grows.

\noindent \textbf{Network Scalability.} In addition to efficient key management, our architectural design contributes significantly to network scalability. Since biometric templates are encrypted using a lightweight stream cipher and the final authentication result is returned in a compact, non-FHE ciphertext (e.g., encrypted using a symmetric cipher), the size of both transmission components remains effectively constant regardless of the number of clients. As the system scales, neither the biometric template nor the authentication result introduces additional ciphertext expansion, making our framework well-suited for seamless deployment in large-scale biometric authentication systems.

%% file: Tabs/key_manage.tex
\begin{table}[htb!]
\caption{Locally stored keys for each party across three FHE-based frameworks for biometric authentication.}
\begin{center}

\begin{tabular}{c c c}
\toprule
\textbf{Party} & \textbf{Framework} & \textbf{Stored Keys} \\
\midrule
\multirow{3}{*}{$\mathcal{TP}$} & $\mathsf{ST-FHE}$ & N/A \\
 & $\mathsf{Orig-TC}$ & N/A \\
 & $\mathsf{Ours}$ & $\mathsf{sk}$, $\mathsf{evk}$ \\

\midrule
\multirow{3}{*}{$\mathcal{C}$} & $\mathsf{ST-FHE}$ & $\{\mathsf{sk_i}\}_{i \in [n_c]}$ \\
 & $\mathsf{Orig-TC}$ & $\{\mathsf{sk_i}\}_{i \in [n_c]}$, $\{\mathbf{k}_i\}_{i \in [n_c]}$ \\
 & $\mathsf{Ours}$ & $\{\mathbf{k}_i\}_{i \in [n_c]}$ \\

\midrule
\multirow{3}{*}{$\mathcal{S}$} & $\mathsf{ST-FHE}$ & $\{\mathsf{evk}_i\}_{i\in [n_c]}$ \\
 & $\mathsf{Orig-TC}$ & $\{\mathsf{evk}_i\}_{i\in [n_c]}$, $\{\mathsf{pk}_{\mathbf{c}_i}\}_{i\in [n_c]}$, $\{\mathsf{dk}_i\}_{i\in [n_c]}$\\
 & $\mathsf{Ours}$ & $\mathsf{evk}$, $\mathsf{pk}_{\mathbf{c}}$, $\{\mathsf{dk}_i\}_{i\in [n_c]}$ \\

\bottomrule

\end{tabular}

\end{center}
\label{tab:key_by_party}
\end{table}

%% file: Sections/07_security_privacy.tex
\section{Security and Privacy of Our Model}
\label{sec:snp}

\subsection{Double Encryption of Symmetric Cipher Key}

To protect the client’s private biometric data from being exposed to third parties, our model assumes two honest-but-curious entities: the server and the trusted party. Both entities follow the protocol but may attempt to learn sensitive information if possible. To prevent either party from independently recovering the stream cipher key \(\mathbf{k}\), we adopt a \textbf{double encryption} strategy.

First, the client encrypts \(\mathbf{k}\) using the FHE scheme under the public key \(\mathsf{pk}_{\mathbf{k}}\), producing the homomorphic decryption key \(\mathsf{dk} = \mathsf{Enc}(\mathbf{k}, \mathsf{pk}_{\mathbf{k}})\). If this ciphertext were sent directly to the server, the trusted party—holding the FHE secret key \(\mathsf{sk}\)—could decrypt \(\mathsf{dk}\) and recover \(\mathbf{k}\), enabling it to derive the keystream and decrypt the client’s biometric data. This would violate the privacy guarantees of the system.

To mitigate this, the client encrypts \(\mathsf{dk}\) again using a symmetric stream cipher with a randomly chosen session key \(\mathbf{k'}\), resulting in \(\mathsf{E}(\mathsf{dk}, \mathbf{k'})\). This ensures that neither the server (which only knows \(\mathbf{k'}\)) nor the trusted party (which only knows \(\mathsf{sk}\)) can independently decrypt and access \(\mathbf{k}\). This design guarantees that \(\mathbf{k}\) remains hidden unless the server and trusted party collude, thereby preserving client privacy under the non-collusion assumption.

\subsection{Trusted Party Authentication}

In our protocol, \(\mathcal{TP}\) enhances result integrity by decrypting the final authentication result \( \mathsf{Enc}(r) \), which is obtained by the server after executing the error tolerance matching algorithm. In contrast, traditional models such as \textsf{ST-FHE} and \textsf{Orig-TC} rely on the client to decrypt and interpret the authentication result, which introduces a critical vulnerability: a compromised client could falsely claim authentication success. Since the server does not possess the FHE secret key \(\mathsf{sk}\), it is unable to decrypt the result on its own and must fully trust the client’s response—thereby opening the door to dishonest behavior and unverifiable outcomes.

By delegating decryption to \(\mathcal{TP}\), our model ensures that the result \(r\) is independently verified and communicated to both the client and the server. This design not only prevents tampering but also eliminates reliance on client honesty, thereby significantly enhancing the integrity and trustworthiness of the authentication process.

%% file: Sections/08_evaluation.tex
\section{Evaluation}
\label{sec:eval}

\subsection{Environment Setup}
\noindent\textbf{System Configuration.} The experiments were conducted on a system powered by a 13th Gen Intel Core i9-13900K processor with 24 cores and 32 threads, running Ubuntu 24.04 LTS. For GPU-accelerated parallel processing, the system was equipped with an NVIDIA GeForce RTX 4060 Ti GPU with 16 GB of memory, utilizing CUDA version 12.4.

\noindent\textbf{Encryption Schemes and Parameters.} The TFHE scheme was employed for all FHE operations, utilizing version 1.1 of the TFHE library to implement the \textsf{BTF} framework. For the TFHE parameters, we selected the $\mathsf{TFHE128}$ and $\mathsf{TFHE80}$ parameter sets as described in ~\cite{tfhe-2}. The Trivium encryption scheme was configured with a fixed security level of 80 bits.

\noindent \textbf{Biometric Feature Size.} In our experiment, we assumed that the biometric data is iris data, where a typical iris scan ($640\times480$ grayscale image) generates approximately 307.2 KB of uncompressed raw data. Additionally, we applied Daugman’s algorithm~\cite{daugman} to extract iris features, resulting in a feature size of 2,048 bits (i.e., 0.25 KB).

\subsection{Comparison of Network Efficiency}

We compared the key sizes during transmission in the network and calculated the total network load for our model, the standard FHE model ($\mathsf{ST-FHE}$), and the original transciphering model ($\mathsf{Orig-TC}$). 

\input{Tabs/keys_table}

\noindent \textbf{Comparison of FHE and Stream Cipher Key Sizes.} Tab.~\ref{tab:tfhe_params} presents the key sizes for both FHE and stream cipher components under 128-bit and 80-bit security levels. Among these, the evaluation key $\mathsf{evk}$ stands out as the largest, reaching 41.6 MB under $\mathsf{TFHE128}$, and remains the primary contributor to client-to-server transmission overhead. In comparison, the public key $\mathsf{pk}_{\mathbf{c}}$, used for encrypting stream cipher ciphertexts $\mathbf{c}$, amounts to 4.93 MB—roughly one-eighth the size of $\mathsf{evk}$. Stream cipher keys such as $\mathbf{k}$ and $\mathbf{k'}$ are notably lightweight, each less than 0.2 MB, making their overhead negligible in practice.

\input{Tabs/tab_eval_kdp}

\noindent\textbf{Significant Network Efficiency in Registration and Verification Stages.} 
A key advantage of our model lies in its minimal communication overhead during the registration and verification phases. For example, a typical iris feature vector extracted using Daugman’s algorithm~\cite{daugman} is approximately 0.25 KB. Under the \(\mathsf{TFHE128}\) parameter set, encrypting this vector homomorphically expands it to approximately 4.93 MB. In contrast, our model transmits the same feature vector as a 0.5 KB stream cipher ciphertext, achieving nearly a \(9,860\times\) reduction in ciphertext size.

The efficiency extends to the authentication result as well. In both the \(\mathsf{ST\text{-}FHE}\) and \(\mathsf{Orig\text{-}TC}\) models, the result \(r \in \{0,1\}\) is returned as a FHE ciphertext, requiring approximately 2.46 KB. By contrast, in our model, the trusted party returns the result as a single-bit symmetric ciphertext, reducing the size by over \(20,000\times\). Although this return ciphertext is relatively small in isolation, its cumulative cost becomes non-negligible in large-scale deployments with frequent authentication events.

\noindent \textbf{Reduced Overhead on the Client-to-Server Channel.} A key advantage of our model during the setup phase is the significant reduction in client-to-server transmission overhead, particularly for large cryptographic keys. In the \(\mathsf{ST\text{-}FHE}\) model, the client must transmit the evaluation key \(\mathsf{evk}\), which alone accounts for 41.6 MB. In the \(\mathsf{Orig\text{-}TC}\) model, the total transmission reaches 46.72 MB, including \(\mathsf{evk}\), \(\mathsf{pk}_{\mathbf{c}}\), \(\mathsf{dk}\), and \(\mathbf{IV}\). By contrast, our model offloads this burden to \(\mathcal{TP}\), which sends the larger keys over a high-bandwidth, low-latency channel. The client transmits only \(\mathbf{k'}\) and \(\mathsf{E}(\mathsf{dk}, \mathbf{k'})\), totaling 394.39~KB. This leads to a significant improvement in network efficiency, achieving approximately \(108\times\) and \(121\times\) reductions in transmission size compared to \(\mathsf{ST\text{-}FHE}\) and \(\mathsf{Orig\text{-}TC}\), respectively. 
% These figures reflect the transmission cost for a single client. As the number of clients increases, the bandwidth savings become even more significant, making our framework highly scalable and efficient in multi-client deployments.

\noindent \textbf{Public Key Overhead in Stream Cipher Encryption.} As noted in Section~\ref{sec:problem}, both the $\mathsf{Orig\text{-}TC}$ model and our model require the server to hold the public key $\mathsf{pk}_{\mathbf{c}}$ for homomorphic decryption of the stream ciphertext. This key comprises LWE samples used to encrypt the extracted biometric feature vector. In our experimental setting, encrypting a 0.25 KB iris feature requires 2,048 LWE samples, which, under the $\mathsf{TFHE128}$ parameter set, results in a total size of 4.93 MB (see Tab.~\ref{tab:kdp}). Although prior work on transciphering often emphasizes the small size of the stream ciphertext $\mathbf{c}$, the overhead associated with transmitting $\mathsf{pk}_{\mathbf{c}}$ is frequently overlooked.

In our model, this transmission is offloaded to the high-bandwidth $\mathcal{TP}$-to-$\mathcal{S}$ channel, avoiding the bottleneck-prone $\mathcal{C}$-to-$\mathcal{S}$ channel used in $\mathsf{Orig\text{-}TC}$. Moreover, $\mathsf{pk}_{\mathbf{c}}$ is transmitted only once during the setup phase and can be reused until the FHE keys are refreshed by the trusted party. As the system scales to $n_c$ clients, the overhead of transmitting $n_c$ distinct public keys $\{\mathsf{pk}_{\mathbf{c}_i}\}_{i \in [n_c]}$ in the $\mathsf{Orig\text{-}TC}$ model becomes increasingly burdensome. In contrast, our model requires only a single, temporarily shared $\mathsf{pk}_{\mathbf{c}}$ generated by $\mathcal{TP}$, offering substantial scalability advantages.

\noindent \textbf{Minor Overhead from Supporting Keys \(\mathsf{pk}_{\mathbf{k}}\) and \(\mathbf{k'}\).} As discussed in Section~\ref{subsec:minor_overhead}, the key \(\mathsf{pk}_{\mathbf{k}}\) is required for the client to encrypt the stream cipher key \(\mathbf{k}\), producing the homomorphic decryption key \(\mathsf{dk}\). In our model, \(\mathsf{pk}_{\mathbf{k}}\) is generated by the trusted party and sent to the client over the $\mathcal{TP}$-to-$\mathcal{C}$ channel. By contrast, this key is locally generated in $\mathsf{Orig\text{-}TC}$ and entirely absent in $\mathsf{ST\text{-}FHE}$. Although \(\mathsf{pk}_{\mathbf{k}}\) introduces a small overhead of 197.19~KB, it is transmitted only once during setup and used solely for encrypting the stream cipher key.

Similarly, the session key \(\mathbf{k'}\) introduces a minor overhead, as it is used to add an additional encryption layer for protecting $\mathsf{dk}$ from the trusted party. This key is of the same size as \(\mathsf{dk}\) (197.19~KB) and is transmitted over the $\mathcal{C}$-to-$\mathcal{S}$ channel. Like \(\mathsf{pk}_{\mathbf{k}}\), the key \(\mathbf{k'}\) is used only during the setup phase, and its contribution to overall communication overhead is minimal.

\subsection{Comparison of Time Performance}

\input{Tabs/tab_su}

\input{Tabs/tab_rs}

\noindent \textbf{Time Performance Comparison with $\mathsf{ST\text{-}FHE}$.} Compared to $\mathsf{ST\text{-}FHE}$, our model significantly reduces the client-side setup time (see Table~\ref{tab:combined_time_complexity}). While our model introduces two operations—double encryption of the symmetric key \(\mathbf{k}\) and stream cipher initialization—these are lightweight. Encrypting the homomorphic decryption key \(\mathsf{dk}\) adds only 5.097 ms, and stream cipher initialization requires approximately 1.01 ms. In contrast, $\mathsf{ST\text{-}FHE}$ demands approximately 0.407 seconds for client-side FHE key generation alone. Since our model outsources FHE key generation to \(\mathcal{TP}\), this overhead is entirely removed from the client, resulting in a faster setup phase.

On the server side, the $\mathsf{ST\text{-}FHE}$ model imposes no computational load during the setup stage. In contrast, our model introduces several setup-phase operations. First, the server decrypts \(\mathsf{E}(\mathsf{dk}, \mathbf{k'})\) using the temporary session key \(\mathbf{k'}\), which takes approximately 4.97 ms as part of the \textsf{KDP}. The dominant time costs stem from initializing the homomorphic stream cipher and generating FHE keystreams, which require 127.291 seconds and 632.476 seconds, respectively. In total, these operations amount to approximately 770.768 seconds when generating \(l_{\mathbf{w}} = 2048\) keystreams. While the FHE keystream generation time is substantial, it occurs only once during setup and can be precomputed ahead of registration or verification, thus not affecting runtime performance during user interaction.

As for the registration and verification stages (see Table~\ref{tab:time_rs_vs}), our model introduces additional requirements compared to $\mathsf{ST-FHE}$ on the server side. Specifically, the server needs to encrypt the stream cipher template under FHE using \(\mathsf{pk}_{\mathbf{c}}\), which takes approximately 33 ms. This overhead is negligible compared to the homomorphic decryption of the stream cipher template, which takes about 3.030 seconds and represents the most significant additional computational cost during this stage. In the verification stage, the decryption of the authentication result \(r\) by the \(\mathcal{TP}\) requires only 98.8 \(\mu\)s. This offsets the need for client-side decryption, reducing its computational burden.

\noindent \textbf{Time Performance Comparison with \(\mathsf{Orig\text{-}TC}\).} Compared to the $\mathsf{Orig\text{-}TC}$ model, our framework introduces an additional encryption layer for the FHE-encrypted symmetric key \(\mathsf{dk}\), which takes approximately 4.96 ms on the client side. However, as in the $\mathsf{ST\text{-}FHE}$ model, our design offloads the FHE key generation process \(\mathsf{Enc.KeyGen}\) to $\mathcal{TP}$. This eliminates the need for the client to perform local key generation, which takes 0.407 seconds in the $\mathsf{Orig\text{-}TC}$ model.

Both $\mathsf{Orig-TC}$ and our model require the initialization of the stream cipher and its FHE counterpart, incurring an additional 770.768 seconds during the setup stage. On the server side, however, our model requires the decryption of \(\mathsf{E}(\mathsf{dk}, \mathbf{k'})\) using the temporary key \(\mathbf{k'}\), which adds a minor overhead of 4.97 ms as part of the \textsf{KDP}.

During the registration and verification stages, both $\mathsf{Orig\text{-}TC}$ and our model perform encryption, $\mathsf{Enc}(\mathbf{c}, \mathsf{pk}_{\mathbf{c}})$, followed by the homomorphic stream cipher decryption, $\mathsf{Eval}(\mathsf{E}^{-1})$. According to $\mathsf{TFHE128}$, these operations take approximately 33 ms and 3.030 seconds, respectively. However, in the $\mathsf{Orig\text{-}TC}$ model, the decryption of the final authentication result \(r\) is handled by the client, which takes approximately \( 0.988\,\mu\text{s} \). In contrast, our model delegates this task to \(\mathcal{TP}\), thereby removing client-side responsibility for result verification.

%% file: Tabs/keys_table.tex
\begin{table}[!htb]
  \centering
  \footnotesize
  \caption{TFHE and Trivium key sizes under 128-bit and 80-bit security levels. All values are in KB, except for $\mathsf{pk}_{\mathbf{c}}$ and $\mathsf{evk}$ in MB.}
  \label{tab:tfhe_params}
  \resizebox{\linewidth}{!}{%
  \begin{tabularx}{\linewidth}{cccccccc}
    \toprule
    \textbf{Param Set} & $\mathsf{sk}$ & $\mathsf{pk}_{\mathbf{k}}$ & $\mathsf{pk}_{\mathbf{c}}$  & $\mathsf{evk}$ & $\mathbf{k}$ & $\mathbf{k'}$ & $\mathsf{dk}$ \\ 
    \midrule
    $\mathsf{TFHE128}$ & 2.46 & 197 & 4.93 MB & \textbf{41.6 MB} & 0.01 & 197 & 197 \\
    $\mathsf{TFHE80}$  & 1.95 & 157  & 3.91 MB & \textbf{23.6 MB} & 0.01 & 157 & 157 \\
    \bottomrule
  \end{tabularx}
  }
\end{table}

%% file: Tabs/tab_eval_kdp.tex
\begin{table*}[htb!]
% \caption{Data transmission during \textsf{KDP}. Data sizes are in KB unless noted (MB).}
\caption{Data transmission during the \textbf{Setup Stage} (\textsf{KDP}, \textsf{INP}). Data sizes are in KB unless otherwise noted. Key sizes are measured according to the $\mathsf{TFHE128}$ parameter set.}
\begin{center}
\begin{tabular}{c c cccccccc}
\toprule
\multicolumn{10}{c}{\textbf{Key Distribution Phase} (\textsf{KDP})} \\
\midrule
\textbf{Framework} & \textbf{Data} & $\mathsf{pk}_\mathbf{k}$ & $\mathsf{pk}_\mathbf{c}$ & $\mathsf{evk}$ & $\mathsf{dk}$ & $\mathbf{IV}$ & $\mathbf{k'}$ & $\mathsf{E}(\mathsf{dk}, \mathbf{k'})$ & \textbf{Total} \\
\midrule
\multirow{3}{*}{$\mathsf{Ours}$} & $\mathcal{TP} \rightarrow \mathcal{C}$ &  197.19 & \xmark & \xmark & \xmark & \xmark & \xmark & \xmark & 197.19 \\
 & $\mathcal{TP} \rightarrow \mathcal{S}$ &  \xmark & 4.93 (MB) & 41.6 (MB) & \xmark & \xmark & \xmark & \xmark & 46.53 (MB) \\
 & $\mathcal{C} \rightarrow \mathcal{S}$ &  \xmark & \xmark & \xmark & \xmark & 0.01 & 197.19 & 197.19 & \fcolorbox{blue}{white}{394.39} \\

\midrule

$\mathsf{ST-FHE}$ & $\mathcal{C} \rightarrow \mathcal{S}$ &  \xmark & \xmark & 41.6 (MB) & \xmark & \xmark & \xmark & \xmark & \fcolorbox{red}{white}{41.6 (MB)} \\

$\mathsf{Orig-TC}$ & $\mathcal{C} \rightarrow \mathcal{S}$ &  \xmark & 4.93 (MB) & 41.6 (MB) & 197.19 & 0.01 & \xmark & \xmark & \fcolorbox{red}{white}{46.72 (MB)} \\
\bottomrule
\end{tabular}

\end{center}
\label{tab:kdp}
\end{table*}

%% file: Tabs/tab_su.tex
\begin{table*}[htb!]
\caption[Time Performance of Setup Stage]{Time Performance of \textbf{Setup Stage} (KDP, INP) in seconds. Colored cells indicate the responsible party: yellow (\(\mathcal{C}\)), green (\(\mathcal{TP}\)), and orange (\(\mathcal{S}\)). ($\mathsf{E^{FHE}.KeyStream}$ generates $l_\mathbf{w} = 2,048$ FHE key stream.)}
\label{tab:combined_time_complexity}

\begin{center}
% \scalebox{1}{
\begin{tabular}{cccccc}
\toprule

\multirow{2}{*}{\textbf{Param Set}} & \multicolumn{5}{c}{\textbf{Key Distribution Phase}} \\
\cmidrule(lr){2-6} 
 & $\mathsf{Enc.KeyGen}$ & $\mathsf{Enc}(\mathbf{k}, \mathsf{pk}_{\mathbf{k}})$ & $\mathsf{E}(\mathsf{dk}, \mathbf{k'})$ & $\mathsf{E}^{-1}(\mathsf{E}(\mathsf{dk}, \mathbf{k'}), \mathbf{k'})$ & \textbf{Total Time} \\

\midrule
$\mathsf{TFHE80}$ & \cellcolor{green!30} \textbf{0.277} & \cellcolor{yellow!60} \num{1.33e-4} & \cellcolor{yellow!60} \num{4.01e-3} & \cellcolor{orange!50} \num{4.07e-3} & 0.287 \\
$\mathsf{TFHE128}$ & \cellcolor{green!30} \textbf{0.407} & \cellcolor{yellow!60} \num{1.37e-4} & \cellcolor{yellow!60} \num{4.96e-3} & \cellcolor{orange!50} \num{4.97e-3} & 0.423 \\

\midrule
$\mathsf{Orig-TC}$ & \cellcolor{yellow!60} \cmark & \cellcolor{yellow!60} \cmark & \xmark & \xmark &  0.407 \\
$\mathsf{ST-FHE}$ & \cellcolor{yellow!60} \cmark & \xmark & \xmark & \xmark &  0.407 \\

\midrule
\midrule

\multirow{2}{*}{\textbf{Param Set}} & \multicolumn{5}{c}{\textbf{Initialization Phase}} \\
\cmidrule(lr){2-6} 
 & $\mathsf{E.Init}$ & $\mathsf{Enc}(\mathbf{IV})$ & $\mathsf{E^{FHE}.Init}$ & $\mathsf{E^{FHE}.KeyStream}$ & \textbf{Total Time} \\

\midrule
$\mathsf{TFHE80}$ & \cellcolor{yellow!60} \num{1.12e-3} & \cellcolor{orange!50} \num{1.32e-4} & \cellcolor{orange!50} 80.179 &  \cellcolor{orange!50} \textbf{399.078} & 479.258 \\
$\mathsf{TFHE128}$ & \cellcolor{yellow!60} \num{1.01e-3} & \cellcolor{orange!50} \num{1.36e-4} & \cellcolor{orange!50} 127.291 & \cellcolor{orange!50} \textbf{643.476} & 770.768 \\

\midrule
$\mathsf{Orig-TC}$ & \cellcolor{yellow!60} \cmark & \cellcolor{orange!50} \cmark & \cellcolor{orange!50} \cmark & \cellcolor{orange!50} \cmark &  770.768 \\
$\mathsf{ST-FHE}$ & \xmark & \xmark & \xmark & \xmark &  0 \\

\bottomrule
\end{tabular}
% }
\end{center}
\end{table*}

%% file: Tabs/tab_rs.tex
\begin{table*}[htb!]
\caption[Time Performance of Registration/Verification Stage]{Time Performance of \textbf{Registration/Verification Stage} in seconds. Colored cells indicate the responsible party: yellow (\(\mathcal{C}\)), green (\(\mathcal{TP}\)), and orange (\(\mathcal{S}\)). D.A. refers to the time depending on the biometric authentication algorithm.}
\label{tab:time_rs_vs}
\begin{center}
% \scalebox{0.7}{
\begin{tabular}{c cccc |cc| c}
\toprule

\multirow{2}{*}{\textbf{Param Set}} & \multicolumn{4}{c}{\textbf{Registration Stage}} & \multicolumn{2}{c}{\textbf{Verification Stage}}\\
\cmidrule(lr){2-8} 
 & $\mathsf{E.KeyStream}$ & $\mathsf{E}(\mathbf{w}, \mathbf{\bar{k}})$ & $\mathsf{Enc}(\mathbf{c}, \mathsf{pk}_{\mathbf{c}})$ & $\mathsf{Eval}(\mathsf{E}^{-1}))$ & $\mathsf{Eval}(\mathsf{ETM}, \cdot, \mathsf{evk})$ & $\mathsf{Dec}(\mathsf{Enc}(r), \mathsf{sk})$ & \textbf{Shared Time} \\

\midrule
$\mathsf{TFHE80}$ & \cellcolor{yellow!60} \num{4.39e-3} & \cellcolor{yellow!60} \num{1.50e-4} & \cellcolor{orange!50} \num{2.63e-2} & \cellcolor{orange!50} \textbf{1.848} &  \cellcolor{orange!50} D.A. & \cellcolor{green!30} \num{7.24e-7} & 1.878 \\
$\mathsf{TFHE128}$ & \cellcolor{yellow!60} \num{4.38e-3} & \cellcolor{yellow!60} \num{1.94e-4} & \cellcolor{orange!50} \num{3.31e-2} & \cellcolor{orange!50} \textbf{3.030} &  \cellcolor{orange!50} D.A. & \cellcolor{green!30} \num{9.88e-7} & 3.067 \\

\midrule
$\mathsf{Orig-TC}$ & \cellcolor{yellow!60} \cmark & \cellcolor{yellow!60} \cmark & \cellcolor{orange!50} \cmark & \cellcolor{orange!50} \cmark &  \cellcolor{orange!50} D.A. &  \cellcolor{yellow!60} \cmark & 3.067 \\
% $\mathsf{ST-FHE}$ & \xmark & \cellcolor{yellow!60} \cmark & \xmark & \xmark &  \cellcolor{orange!50} D.A. &  \cellcolor{yellow!60} \cmark & \num{1.94e-4} \\
\bottomrule
\end{tabular}
% }
\end{center}
\end{table*}

%% file: Sections/09_conclusion.tex
% \section{Conclusion}

% In this paper, we presented a novel privacy-preserving bidirectional transciphering framework, \textsf{BTF}, which effectively minimizes the network overhead associated with FHE ciphertext in existing FHE-based biometric systems. The \textsf{BTF} model employs double encryption of the stream cipher key, providing security not only against external attackers but also against semi-honest parties such as the server and the trusted party. This approach combines the efficiency of lightweight symmetric encryption during both the transmission and return phases with the privacy-preserving properties of FHE for secure computation. Moreover, the final result can only be decrypted by the trusted party, further enhancing the security of the model.

\section{Conclusion}

We presented \textsf{BTF}, a bidirectional transciphering framework for privacy-preserving biometric authentication that significantly reduces the network overhead of FHE-based systems. Through a double encryption mechanism, \textsf{BTF} protects sensitive data from semi-honest parties under a non-collusion assumption. By leveraging lightweight symmetric encryption during transmission and return, and delegating decryption to a trusted party, \textsf{BTF} improves both efficiency and result integrity. Our implementation and evaluation confirm its practicality and scalability for real-world deployments. Beyond authentication, \textsf{BTF} can serve as a general bidirectional framework for FHE-based applications that seek to minimize network overhead.